\input harvmac

\input amssym.def
\input amssym.tex
\noblackbox

\def\ZZ{{\Bbb Z}}
\def\RR{{\Bbb R}}
\def\CC{{\Bbb C}}

\def\HH{{\Bbb H}}

\def\PP{{\Bbb P}}

\def\A{{\cal A}}
\def\B{{\cal B}}
\def\D{{\cal D}}
\def\N{{\cal N}}
\def\G{{\cal G}}
\def\F{{\cal F}}
\def\H{{\cal H}}

\def\M{{\cal M}}

%\HarnadHH
\lref\HarnadHH{
J.~Harnad and J.~McKay,
%``Modular Solutions to Equations of Generalized Halphen Type,''
Proc.\ Roy.\ Soc.\ Lond.\  {\bf 456}, 261 (2000)
[arXiv:solv-int/9804006].
%%CITATION = SOLV-INT 9804006;%%
}

%\EdelsteinSP
\lref\EdelsteinSP{
J.~D.~Edelstein, M.~Marino and J.~Mas,
%``Whitham hierarchies, instanton corrections and soft supersymmetry  breaking
%in N = 2 SU(N) super Yang-Mills theory,''
Nucl.\ Phys.\ B {\bf 541}, 671 (1999)
[arXiv:hep-th/9805172].
%%CITATION = HEP-TH 9805172;%%
J.~D.~Edelstein, M.~Gomez-Reino and J.~Mas,
%``Instanton corrections in N = 2 supersymmetric theories with classical  gauge
%groups and fundamental matter hypermultiplets,''
Nucl.\ Phys.\ B {\bf 561}, 273 (1999)
[arXiv:hep-th/9904087].
%%CITATION = HEP-TH 9904087;%%
J.~D.~Edelstein, M.~Gomez-Reino, M.~Marino and J.~Mas,
%``N = 2 supersymmetric gauge theories with massive hypermultiplets and  the
%Whitham hierarchy,''
Nucl.\ Phys.\ B {\bf 574}, 587 (2000)
[arXiv:hep-th/9911115].
%%CITATION = HEP-TH 9911115;%%
}

%\MarinoGJ
\lref\MarinoGJ{
M.~Marino and G.~W.~Moore,
%``Integrating over the Coulomb branch in N = 2 gauge theory,''
Nucl.\ Phys.\ Proc.\ Suppl.\  {\bf 68}, 336 (1998)
[arXiv:hep-th/9712062].
%%CITATION = HEP-TH 9712062;%%
M.~Marino and G.~W.~Moore,
%``The Donaldson-Witten function for gauge groups of rank larger than one,''
Commun.\ Math.\ Phys.\  {\bf 199}, 25 (1998)
[arXiv:hep-th/9802185].
%%CITATION = HEP-TH 9802185;%%
M.~Marino,
%``The uses of Whitham hierarchies,''
Prog.\ Theor.\ Phys.\ Suppl.\  {\bf 135}, 29 (1999)
[arXiv:hep-th/9905053].
%%CITATION = HEP-TH 9905053;%%
J.~D.~Edelstein, M.~Gomez-Reino and M.~Marino,
%``Blowup formulae in Donaldson-Witten theory and integrable hierarchies,''
Adv.\ Theor.\ Math.\ Phys.\  {\bf 4}, 503 (2000)
[arXiv:hep-th/0006113].
%%CITATION = HEP-TH 0006113;%%
J.~D.~Edelstein, M.~Gomez-Reino and M.~Marino,
%``Remarks on twisted theories with matter,''
JHEP {\bf 0101}, 004 (2001)
[arXiv:hep-th/0011227].
%%CITATION = HEP-TH 0011227;%%
J.~D.~Edelstein and M.~Gomez-Reino,
%``Integrable hierarchies in Donaldson-Witten and Seiberg-Witten theories,''
arXiv:hep-th/0010061.
%%CITATION = HEP-TH 0010061;%%
}

%\ManinWJ
\lref\ManinWJ{ Y.~I.~Manin and P.~Zograf,
%``Invertible cohomological field theories and Weil-Petersson volumes,''
arXiv:math.ag/9902051.
%%CITATION = MATH-AG 9902051;%%
}

%\ZografUR
\lref\ZografUR{ P.~Zograf,
%``Weil-Petersson volumes of moduli spaces of curves and the genus expansion in
%two dimensional gravity,''
arXiv:math.ag/9811026.
%%CITATION = MATH-AG 9811026;%%
}

%\GopakumarJQ
\lref\GopakumarJQ{
R.~Gopakumar and C.~Vafa,
%``M-theory and topological strings. I,''
arXiv:hep-th/9809187;
%%CITATION = HEP-TH 9809187;%%
%``M-theory and topological strings. II,''
arXiv:hep-th/9812127.
%%CITATION = HEP-TH 9812127;%%
%\GopakumarII
}

%%\MatoneYR
\lref\MatoneYR{ M.~Matone,
%``The Higgs model For Anyons And
%Liouville Action: Chaotic Spectrum, Energy Gap And Exclusion
%Principle,''
Mod.\ Phys.\ Lett.\ A {\bf 9}, 1673 (1994)
[arXiv:hep-th/9403079];
%%%CITATION = HEP-TH 9403079;%%
%}
%%\MatoneJQ
%\lref\MatoneJQ{M.~Matone,
%``The Higgs model For Anyons And Liouville Action,''
J.\ Geom.\ Phys.\ {\bf 17}, 49 (1995) [arXiv:hep-th/9310149].
%%%CITATION = HEP-TH 9310149;%%
}

%\MatoneUY
\lref\MatoneUY{ M.~Matone,
%``Eigenfunctions of the Laplacian Acting on Degree Zero Bundles over Special
%Riemann Surfaces,''
Trans.\ Am.\ Math.\ Soc.\ {\bf 356}, 2989 (2004)
[arXiv:math.ag/0105051].
%%CITATION = MATH-AG 0105051;%%
}

%\MatoneNT
\lref\MatoneNT{ M.~Matone,
%``The instanton universal moduli space
%of N = 2 supersymmetric Yang-Mills theory,''
Phys.\ Lett.\ B {\bf
514}, 161 (2001) [arXiv:hep-th/0105041].
%%CITATION = HEP-TH 0105041;%%
}

%\MatoneWN
\lref\MatoneWN{ M.~Matone,
%``Multi-instanton measure from recursion relations in N = 2 supersymmetric
%Yang-Mills theory,''
JHEP {\bf 0104}, 041 (2001) [arXiv:hep-th/0103246].
%%CITATION = HEP-TH 0103246;%%
}

%\BonelliRY
\lref\BonelliRY{ G.~Bonelli, M.~Matone and M.~Tonin,
%``Solving N = 2 SYM by reflection symmetry of quantum vacua,''
Phys.\ Rev.\ D {\bf 55}, 6466 (1997) [arXiv:hep-th/9610026].
%%CITATION = HEP-TH 9610026;%%
%}
%\FlumeIX
%\lref\FlumeIX{
R.~Flume, M.~Magro, L.~O'Raifeartaigh, I.~Sachs and
O.~Schnetz,
%``Uniqueness of the Seiberg-Witten effective Lagrangian,''
Nucl.\ Phys.\ B {\bf 494}, 331 (1997) [arXiv:hep-th/9611123].
%%CITATION = HEP-TH 9611123;%%
}

%\BonelliQH
\lref\BonelliQH{ G.~Bonelli and M.~Matone,
%``Nonperturbative Relations in N=2 Susy Yang-Mills and WDVV Equation,''
Phys.\ Rev.\ Lett.\ {\bf 77}, 4712 (1996) [arXiv:hep-th/9605090].
%%CITATION = HEP-TH 9605090;%%
%\BertoldiDK
G.~Bertoldi and M.~Matone,
%\BertoldiAE
%``N = 2 SYM RG scale as modulus for WDVV equations,''
Phys.\ Rev.\ D {\bf 57}, 6483 (1998)
[arXiv:hep-th/9712109].
%%CITATION = HEP-TH 9712109;%%
}

%\EynardXT
\lref\EynardXT{ B.~Eynard, A.~Kokotov and D.~Korotkin,
%``Genus one contribution to free energy in hermitian two-matrix model,''
arXiv:hep-th/0403072.
%%CITATION = HEP-TH 0403072;%%
}

%\MatoneKR
\lref\MatoneKR{ M.~Matone,
%``Liouville Equation And Schottky Problem,''
Lett.\ Math.\ Phys.\ {\bf 33}, 75 (1995) [arXiv:hep-th/9310066].
%%CITATION = HEP-TH 9310066;%%
}

%\EguchiUI
\lref\EguchiUI{ T.~Eguchi and H.~Ooguri,
%``Chiral Bosonization On A Riemann Surface,''
Phys.\ Lett.\ B {\bf 187}, 127 (1987).
%%CITATION = PHLTA,B187,127;%%
}

\lref\LTA{L.~A.~Takhtajan, in {\it The Verdier Memorial Conference
on Integrable Systems}, Birkh\"auser, Boston, 1993.
M.~J.~Ablowitz, S.~Chakravarty and R.~G.~Halburd, J.\ Math.\
Phys.\ {\bf 44}, 3147 (2003). }

%\FlumeRP
\lref\FlumeRP{ R.~Flume, F.~Fucito, J.~F.~Morales and
R.~Poghossian,
%``Matone's relation in the presence of gravitational couplings,''
JHEP {\bf 0404}, 008 (2004) [arXiv:hep-th/0403057].
%%CITATION = HEP-TH 0403057;%%
}

%\BellisaiBC
\lref\BellisaiBC{ D.~Bellisai, F.~Fucito, A.~Tanzini and
G.~Travaglini,
%``Instanton calculus, topological field theories
%and N = 2 super Yang-Mills theories,''
JHEP {\bf 0007}, 017
(2000) [arXiv:hep-th/0003272].
%%CITATION = HEP-TH 0003272;%%
%``Multi-instantons, supersymmetry and topological field
%theories,''
Phys.\ Lett.\ B {\bf 480}, 365 (2000)
[arXiv:hep-th/0002110].
%%CITATION = HEP-TH 0002110;%%
%}
%\BruzzoDI
%\lref\BruzzoDI{
U.~Bruzzo, F.~Fucito, A.~Tanzini and
G.~Travaglini,
%``On the multi-instanton measure for super Yang-Mills theories,''
Nucl.\ Phys.\ B {\bf 611}, 205 (2001) [arXiv:hep-th/0008225].
%%CITATION = HEP-TH 0008225;%%
%}
%\BruzzoXF
%\lref\BruzzoXF{
U.~Bruzzo, F.~Fucito, J.~F.~Morales and A.~Tanzini,
%``Multi-instanton calculus and equivariant cohomology,''
JHEP {\bf 0305}, 054 (2003) [arXiv:hep-th/0211108].
%%CITATION = HEP-TH 0211108;%%
%}
%\BruzzoRW
%\lref\BruzzoRW{
U.~Bruzzo and F.~Fucito,
%``Superlocalization formulas and supersymmetric Yang-Mills theories,''
Nucl.\ Phys.\ B {\bf 678}, 638 (2004) [arXiv:math-ph/0310036].
%%CITATION = MATH-PH 0310036;%%
}

%\FucitoUA
\lref\FucitoUA{ F.~Fucito and G.~Travaglini,
%``Instanton calculus and nonperturbative relations in N = 2 supersymmetric
%gauge theories,''
Phys.\ Rev.\ D {\bf 55}, 1099 (1997) [arXiv:hep-th/9605215].
%%CITATION = HEP-TH 9605215;%%
}

%\NekrasovRJ
\lref\NekrasovRJ{ N.~Nekrasov and A.~Okounkov,
%``Seiberg-Witten theory and random partitions,''
arXiv:hep-th/0306238.
%%CITATION = HEP-TH 0306238;%%
}

%\NekrasovAF
\lref\NekrasovAF{ N.~A.~Nekrasov,
%``Seiberg-Witten prepotential from instanton counting,''
arXiv:hep-th/0306211.
%%CITATION = HEP-TH 0306211;%%
}

%\KlemmPA
\lref\KlemmPA{ A.~Klemm, M.~Marino and S.~Theisen,
%``Gravitational corrections in supersymmetric gauge theory and matrix models,''
JHEP {\bf 0303}, 051 (2003) [arXiv:hep-th/0211216].
%%CITATION = HEP-TH 0211216;%%
%}
%\DijkgraafYN
%\lref\DijkgraafYN{
R.~Dijkgraaf, A.~Sinkovics and M.~Temurhan,
%``Matrix models and gravitational corrections,''
arXiv:hep-th/0211241.
%%CITATION = HEP-TH 0211241;%%
}

%\NakajimaUH
\lref\NakajimaUH{ H.~Nakajima and K.~Yoshioka,
%``Instanton counting on blowup. I,''
arXiv:math.ag/0306198;
%``Lectures on instanton counting,''
arXiv:math.ag/0311058.
%%CITATION = MATH-AG 0306198;%%
%%CITATION = MATH-AG 0311058;%%
}

\lref\SeibergRS{ N.~Seiberg and
E.~Witten,
% ``Electric - magnetic duality, monopole condensation, and confinement in N=2
%supersymmetric Yang-Mills theory,''
Nucl.\ Phys.\ B {\bf 426}, 19 (1994)
[Erratum-ibid.\ B {\bf 430}, 485 (1994)]
[arXiv:hep-th/9407087];
%%CITATION = HEP-TH 9407087;%%
%``Monopoles, duality and chiral symmetry breaking in N=2 supersymmetric QCD,''
Nucl.\ Phys.\ B {\bf 431}, 484 (1994)
[arXiv:hep-th/9408099].
%%CITATION = HEP-TH 9408099;%%
}

\lref\MatoneRX{
M.~Matone,
%``Instantons and recursion relations in N=2 SUSY gauge theory,''
Phys.\ Lett.\ B {\bf 357}, 342 (1995) [arXiv:hep-th/9506102];
%%CITATION = HEP-TH 9506102;%%
%}
%\MatoneJR
%\lref\MatoneJR{ M.~Matone,
%``Koebe 1/4 theorem and inequalities in N=2 superQCD,''
Phys.\ Rev.\ D {\bf 53}, 7354 (1996) [arXiv:hep-th/9506181].
%%CITATION = HEP-TH 9506181;%%
}

%\CallanEM
\lref\CallanEM{ C.~G.~.~Callan and F.~Wilczek,
%``Infrared Behavior At Negative Curvature,''
 Nucl.\ Phys.\ B {\bf 340}, 366 (1990).
%%CITATION = NUPHA,B340,366;%%
}

%\GrignaniZM
\lref\GrignaniZM{ G.~Grignani, P.~Orland, L.~D.~Paniak and
G.~W.~Semenoff,
%``Matrix theory interpretation of DLCQ string worldsheets,''
Phys.\ Rev.\ Lett.\ {\bf 85}, 3343 (2000) [arXiv:hep-th/0004194].
%%CITATION = HEP-TH 0004194;%%
}

\lref\Kontsevichetal{M.~Kontsevich and A.~Zorich, math.gt/0201292.
%Title: Connected components of the moduli spaces of Abelian
%differentials with prescribed singularities
A.~Eskin, H.~Masur and A.~Zorich, math.ds/0202134.
%Title: Moduli Spaces of
%Abelian Differentials: The Principal Boundary, Counting Problems
%and the Siegel-Veech Constants
A.~Kokotov and D.~Korotkin, math.dg/0405042. }
% math.dg/0405042
%Title: Tau-functions on spaces of holomorphic differentials over
%Riemann surfaces and determinants of Laplacians in flat metrics
%with conic singularities

\lref\MarshakovAE{ A.~Marshakov, A.~Mironov and A.~Morozov,
%``WDVV-like equations in N = 2 SUSY Yang-Mills theory,''
Phys.\ Lett.\ B {\bf 389}, 43 (1996)
[arXiv:hep-th/9607109];
%%CITATION = HEP-TH 9607109;%%
%``More evidence for the WDVV equations in N = 2 SUSY Yang-Mills theories,''
Int.\ J.\ Mod.\ Phys.\ A {\bf 15}, 1157 (2000)
[arXiv:hep-th/9701123].
%%CITATION = HEP-TH 9701123;%%
}
\lref\ItoZR{ K.~Ito and S.~K.~Yang,
%``The WDVV equations in N = 2 supersymmetric Yang-Mills theory,''
Phys.\ Lett.\ B {\bf 433}, 56 (1998)
[arXiv:hep-th/9803126].
%%CITATION = HEP-TH 9803126;%%
}
 \lref\ChanGJ{ G.~Chan and E.~D'Hoker,
%``Instanton recursion relations for the effective prepotential in N = 2 super
%Yang-Mills,''
Nucl.\ Phys.\ B {\bf 564}, 503 (2000)
[arXiv:hep-th/9906193].
%%CITATION = HEP-TH 9906193;%%
}\lref\LosevTP{ A.~Losev, N.~Nekrasov and S.~L.~Shatashvili,
%``Issues in topological gauge theory,''
Nucl.\ Phys.\ B {\bf 534},
549 (1998) [arXiv:hep-th/9711108];
%%CITATION = HEP-TH 9711108;%%
%``Testing Seiberg-Witten solution,''
arXiv:hep-th/9801061.
%%CITATION = HEP-TH 9801061;%%
}\lref\MatoneWH{
M.~Matone,
%``Seiberg-Witten duality in Dijkgraaf-Vafa theory,''
Nucl.\ Phys.\ B {\bf 656}, 78 (2003) [arXiv:hep-th/0212253].
%%CITATION = HEP-TH 0212253;%%
}\lref\MatoneWN{ M.~Matone,
%``Multi-instanton measure from recursion relations in N = 2 supersymmetric
%Yang-Mills theory,''
JHEP {\bf 0104}, 041 (2001)
[arXiv:hep-th/0103246].
%%CITATION = HEP-TH 0103246;%%
}\lref\NakayamaVK{ Y.~Nakayama,
%``Liouville field theory: A decade after the revolution,''
arXiv:hep-th/0402009.
%%CITATION = HEP-TH 0402009;%%
}\lref\GoulianQR{ M.~Goulian and M.~Li,
%``Correlation Functions In Liouville Theory,''
Phys.\ Rev.\ Lett.\ {\bf 66}, 2051 (1991).
%%CITATION = PRLTA,66,2051;%%
}\lref\DornXN{ H.~Dorn and H.~J.~Otto,
%``Two and three point functions in Liouville theory,''
Nucl.\ Phys.\ B {\bf 429}, 375 (1994)
[arXiv:hep-th/9403141].
%%CITATION = HEP-TH 9403141;%%
}\lref\ZamolodchikovAA{
A.~B.~Zamolodchikov and A.~B.~Zamolodchikov,
%``Structure constants and conformal bootstrap in Liouville field theory,''
Nucl.\ Phys.\ B {\bf 477}, 577 (1996)
[arXiv:hep-th/9506136].
%%CITATION = HEP-TH 9506136;%%
}\lref\NakayamaEP{
Y.~Nakayama,
%``Tadpole cancellation in unoriented Liouville theory,''
JHEP {\bf 0311}, 017 (2003)
[arXiv:hep-th/0309063].
%%CITATION = HEP-TH 0309063;%%
}\lref\HoriAX{ K.~Hori and A.~Kapustin,
% ``Duality of the fermionic 2d black hole and N = 2 Liouville theory as mirror symmetry,''
JHEP {\bf 0108}, 045 (2001)
[arXiv:hep-th/0104202].
%%CITATION = HEP-TH 0104202;%%
}\lref\AganagicQJ{
M.~Aganagic, R.~Dijkgraaf, A.~Klemm, M.~Marino and C.~Vafa,
%``Topological strings and integrable hierarchies,''
arXiv:hep-th/0312085.
%%CITATION = HEP-TH 0312085;%%
}\lref\DijkgraafFC{ R.~Dijkgraaf and C.~Vafa,
%``Matrix models,
%topological strings, and supersymmetric gauge theories,''
Nucl.\ Phys.\ B {\bf 644}, 3 (2002) [arXiv:hep-th/0206255].
%%CITATION = HEP-TH 0206255;%%
}\lref\NekrasovQD{
N.~A.~Nekrasov,
%``Seiberg-Witten prepotential from instanton counting,''
arXiv:hep-th/0206161.
%%CITATION = HEP-TH 0206161;%%
}\lref\NekrasovRJ{ N.~Nekrasov and A.~Okounkov,
%``Seiberg-Witten
%theory and random partitions,''
arXiv:hep-th/0306238.
%%CITATION = HEP-TH 0306238;%%
} \lref\DoreyIK{ N.~Dorey, T.~J.~Hollowood, V.~V.~Khoze and
M.~P.~Mattis,
%``The calculus of many instantons,''
Phys.\ Rept.\
{\bf 371}, 231 (2002) [arXiv:hep-th/0206063].
%%CITATION = HEP-TH 0206063;%%
}\lref\CMS{L.~Cantini, P.~Menotti and D.~Seminara,
%``Proof of
%Polyakov conjecture for general elliptic singularities,''
 Phys.\ Lett.\ B {\bf 517}, {203}{(2001)} [arXiv:hep-th/0105081];
% ``Liouville theory, accessory parameters and 2+1 dimensional gravity,''
 Nucl.\ Phys.\ B {\bf 638}, {351}{(2002)} [arXiv:hep-th/0203103].}
\lref\zota{P.~G.~Zograf and L.~A.~Takhtajan,
%``On Liouville equation,
%accessory parameters, and the geometry of Teichm\"uller space for
%Riemann surfaces of genus 0,"
Math.\ USSR\ Sb.\ {\bf 60}, {143} (1988);
%``On the uniformization of Riemann surfaces an on
%the Weil-Petersson metric on the Teichm\"uller and Schottky
%spaces,''
Math.\ USSR\ Sb.\ {\bf 60}, 297 (1988);
%``Hyperbolic 2-spheres
%with conical singularities, accessory parameters and Kaehler
%metrics on M0,n'',
Trans.\ Am.\ Math.\ Soc.\ {\bf 355}, {1857} {(2002)} [arXiv:
math.cv/0112170].
%%CITATION = MATH-CV 0112170;%%
L.~A.~Takhtajan and L.~P.~Teo,
%``Liouville action and Weil-Petersson metric on
%deformation spaces, global Kleinian reciprocity and holography'',
Commun.\ Math.\ Phys.\ {\bf 239}, 183 (2003) [arXiv:
math.cv/0204318];
%%CITATION = MATH-CV 0204318;%%
%\TakhtajanHM
%\lref\TakhtajanHM{ L.~A.~Takhtajan and L.~P.~Teo,
%``Weil-Petersson metric on the universal Teichmuller space I: Curvature
%properties and Chern forms,''
arXiv:math.cv/0312172.
%%CITATION = MATH-CV 0312172;%%
%}
 }\lref\DeligneKnudsenMumford{P.~Deligne and D.~Mumford,
%``On the
%irreducibility of the space of curves of given genus,"
IHES\ Publ.\ Math.\ {\bf 36}, 75 (1969). F.~Knudsen and
D.~Mumford,
%``The
%projectivity of the moduli space of stable curves I: Preliminaries
%on ``det'' and ``div'',"
Math.\ Scand.\ {\bf 39}, 19 (1976). F.~Knudsen,
%``Projectivity of the moduli space of stable curves,
%II, the stacks,"
Math.\ Scand.\ {\bf 52}, 200 (1983).}\lref\Zog{ P.~G.~Zograf,
%``The Weil-Petersson Volume of the Moduli Space of
%Punctured Spheres,"
Cont.\ Math.\ Amer.\ AMS {\bf 150}, 267 (1993).}\lref\wolpertis{
S.~A.~Wolpert,
%``On the homology of the
%moduli space of stable curves,"
Ann.\ of Math.\ {\bf 118}, 491 (1983);
%``On the Weil-Petersson geometry of the moduli space of
%curves"
Amer.\ J.\ Math.\ {\bf 107}, 969 (1985)
.}\lref\MatonePZ{M.~Matone,
%``Nonperturbative model Of Liouville
%Gravity,''
J.\ Geom.\ Phys.\ {\bf 21}, 381 (1997) [arXiv:hep-th/9402081].
%%CITATION = HEP-TH 9402081;%%
}\lref\bomama{G.~Bonelli, P.~A.~Marchetti and M.~Matone,
%``Nonperturbative 2-D Gravity, Punctured Spheres And Theta Vacua In String
%Theories,''
Phys.\ Lett.\ B {\bf 339}, 49 (1994)
[arXiv:hep-th/9407091];
%%CITATION = HEP-TH 9407091;%%
%``Algebraic - geometrical formulation of two-dimensional quantum gravity,''
Lett.\ Math.\ Phys.\ {\bf 36}, 189 (1996) [arXiv:hep-th/9502089].
%%CITATION = HEP-TH 9502089;%%

} \lref\manin{M.~Kontsevich and Y.~I.~Manin (with Appendix by R.
Kaufmann),
%``Quantum cohomology of a product,"
Invent.\ Math.\ {\bf 124}, 313 (1996)  arXiv:q-alg/9502009.
R.~Kaufmann, Y.~I.~Manin and D.~Zagier,
%``Higher
%Weil-Petersson Volumes of Moduli Spaces of Stable n-pointed
%Curves,''
Comm.\ Math.\ Phys.\ {\bf 181}, 763 (1996)
arXiv:alg-geom/9604001.}\lref\Givental{A.~Givental,
%``Semisimple Frobenius
%structures at higher genus'',
math.ag/0008067.}\lref\underground{J.~S.~Song and Y.~S.~Song,
%``Notes from the underground: A propos of Givental's conjecture,''
hep-th/0103254.}

\lref\Yuri{Y.~I.~Manin, ``Frobenius Manifolds, Quantum Cohomology,
and Moduli Spaces," AMS\ Colloquium\ Publications {\bf 47}, 1999.}
%\WittenHR
\lref\WittenHR{
E.~Witten,
%``Two-Dimensional Gravity And Intersection Theory On Moduli Space,''
Surveys Diff.\ Geom.\ {\bf 1}, 243 (1991).
%%CITATION = 00078,1,243;%%
}

\lref\DijkgraafPP{ R.~Dijkgraaf, S.~Gukov, V.~A.~Kazakov and
C.~Vafa,
%``Perturbative analysis of gauged matrix models,''
Phys.\
Rev.\ D {\bf 68}, 045007 (2003) [arXiv:hep-th/0210238].
%%CITATION = HEP-TH 0210238;%%
}

\lref\DoreyZJ{ N.~Dorey, V.~V.~Khoze and M.~P.~Mattis,
%``Multi-instanton check of the relation between the prepotential F and the
%modulus u in N = 2 SUSY Yang-Mills theory,''
Phys.\ Lett.\ B {\bf 390}, 205 (1997)
[arXiv:hep-th/9606199].
%%CITATION = HEP-TH 9606199;%%
}\lref\HowePW{
P.~S.~Howe and P.~C.~West,
%``Superconformal Ward identities and N = 2 Yang-Mills theory,''
Nucl.\ Phys.\ B {\bf 486}, 425 (1997)
[arXiv:hep-th/9607239].
%%CITATION = HEP-TH 9607239;%%
}\lref\IqbalIX{ A.~Iqbal and A.~K.~Kashani-Poor,
%``Instanton counting and Chern-Simons theory,''
Adv.\ Theor.\ Math.\ Phys.\ {\bf 7}, 457 (2004)
[arXiv:hep-th/0212279];
%%CITATION = HEP-TH 0212279;%%
%``SU(N) geometries and topological string amplitudes,''
arXiv:hep-th/0306032.
%%CITATION = HEP-TH 0306032;%%
}\lref\EguchiSJ{
T.~Eguchi and H.~Kanno,
%``Topological strings and Nekrasov's formulas,''
JHEP {\bf 0312}, 006 (2003)
[arXiv:hep-th/0310235].
%%CITATION = HEP-TH 0310235;%%
}\lref\KatzFH{
S.~Katz, A.~Klemm and C.~Vafa,
%``Geometric engineering of quantum field theories,''
Nucl.\ Phys.\ B {\bf 497}, 173 (1997)
[arXiv:hep-th/9609239].
%%CITATION = HEP-TH 9609239;%%
} \lref\AntoniadisZE{ I.~Antoniadis, E.~Gava,
K.~S.~Narain and T.~R.~Taylor,
%``Topological amplitudes in string theory,''
Nucl.\ Phys.\ B {\bf 413}, 162 (1994)
[arXiv:hep-th/9307158].
%%CITATION = HEP-TH 9307158;%%
}\lref\BershadskyCX{ M.~Bershadsky, S.~Cecotti, H.~Ooguri and
C.~Vafa,
%``Kodaira-Spencer theory of gravity and exact results for
%quantum string amplitudes,''
Commun.\ Math.\ Phys.\ {\bf 165}, 311 (1994)
[arXiv:hep-th/9309140].
%%CITATION = HEP-TH 9309140;%%
}\lref\ZamolodchikovYB{ A.~Zamolodchikov,
%``Higher
%equations of motion in Liouville field theory,''
arXiv:hep-th/0312279.
%%CITATION = HEP-TH 0312279;%%
}\lref\PolyakovTJ{ A.~M.~Polyakov,
%``String theory and quark confinement,''
Nucl.\ Phys.\ Proc.\ Suppl.\ {\bf 68}, 1 (1998)
[arXiv:hep-th/9711002].
%%CITATION = HEP-TH 9711002;%%
}
 \lref\AhnSX{ C.~Ahn, C.~Kim, C.~Rim and
M.~Stanishkov,
%``Duality in N = 2 super-Liouville theory,''
arXiv:hep-th/0210208.
%%CITATION = HEP-TH 0210208;%%
}\lref\HoriCD{
K.~Hori and A.~Kapustin,
%``Worldsheet descriptions of wrapped NS five-branes,''
JHEP {\bf 0211}, 038 (2002)
[arXiv:hep-th/0203147].
%%CITATION = HEP-TH 0203147;%%
}\lref\GhoshalWM{ D.~Ghoshal and
C.~Vafa,
%``C = 1 string as the topological theory of the conifold,''
Nucl.\ Phys.\ B {\bf 453}, 121 (1995)
[arXiv:hep-th/9506122].
%%CITATION = HEP-TH 9506122;%%
}\lref\GH{Ph.~Griffiths and J.~Harris, ``Principles of Algebraic
Geometry,'' John Wiley and Sons, 1978.}

\lref\Duistermaat{ {J.~J.~Duistermaat and G.~J.~Heckman,
%``On the variation in the cohomology
%of the symplectic form of the reduced phase space,"
Invent.\ Math.\ {\bf 69}, 259 (1982).}}

\lref\DavidHJ{ F.~David,
%``Conformal Field Theories Coupled To 2-D
%Gravity In The Conformal Gauge,''
Mod.\ Phys.\ Lett.\ A {\bf 3},
1651 (1988).
%%CITATION = MPLAE,A3,1651;%%
J.~Distler and H.~Kawai,
% ``Conformal Field Theory And 2-D Quantum Gravity Or Who's Afraid Of Joseph
%Liouville?,''
Nucl.\ Phys.\ B {\bf 321}, 509 (1989).
%%CITATION = NUPHA,B321,509;%%
}\lref\GrossZZ{
D.~J.~Gross and J.~Walcher,
%``Non-perturbative RR potentials in the c(hat) = 1 matrix model,''
arXiv:hep-th/0312021.
%%CITATION = HEP-TH 0312021;%%
}\lref\KontsevichQZ{ M.~Kontsevich and Y.~I.~Manin,
%``Gromov-Witten Classes, Quantum
%Cohomology, And Enumerative Geometry,''
Commun.\ Math.\ Phys.\
{\bf 164}, 525 (1994) [arXiv:hep-th/9402147].
%%CITATION = HEP-TH 9402147;%%
}\lref\DiFrancescoXR{
P.~Di Francesco and C.~Itzykson,
%``Quantum intersection rings,''
arXiv:hep-th/9412175.
%%CITATION = HEP-TH 9412175;%%
%}
%%\DubrovinHC
%\lref\DubrovinHC{
B.~Dubrovin,
%%``Geometry Of 2-D Topological Field Theories,''
arXiv:hep-th/9407018.
%%%CITATION = HEP-TH 9407018;%%
}
%
%\MatoneTJ
\lref\MatoneTJ{
M.~Matone,
%``Uniformization Theory And 2-D Gravity. 1. Liouville Action And Intersection
%Numbers,''
Int.\ J.\ Mod.\ Phys.\ A {\bf 10}, 289 (1995)
[arXiv:hep-th/9306150].
%%CITATION = HEP-TH 9306150;%%
}

\lref\mm{ E.~Brezin and V.~A.~Kazakov,
%``Exactly Solvable Field
%Theories Of Closed Strings,''
Phys.\ Lett.\ B {\bf 236}, 144
(1990).
%%CITATION = PHLTA,B236,144;%%
 M.~R.~Douglas and S.~H.~Shenker,
%``Strings In Less Than One-Dimension,''
Nucl.\ Phys.\ B {\bf 335},
635 (1990).
%%CITATION = NUPHA,B335,635;%%
D.~J.~Gross and A.~A.~Migdal,
%``Nonperturbative Two-Dimensional Quantum Gravity,''
Phys.\ Rev.\ Lett.\ {\bf 64}, 127 (1990);
%%CITATION = PRLTA,64,127;%%
%``Nonperturbative Solution Of The Ising model On A Random Surface,''
Phys.\ Rev.\ Lett.\ {\bf 64}, 717 (1990). }

\lref\bookss{ Y.~Imayoshi and M.~Taniguchi, ``An introduction to
the Teichm\"uller spaces," Springer-Verlag, 1992. P.~Buser,
``Geometry and spectra of compact Riemann surfaces," Birkh\"auser,
1992.}

\lref\zztop{ P.~G.~Zograf,
%``The Liouville action on moduli spaces,
%and uniformization of degenerating Riemann surfaces,"
Leningrad Math.\ J.\ {\bf 1}, 941 (1990).}

%\KonishiQY
\lref\KonishiQY{
Y.~Konishi and M.~Naka,
%``Geometric engineering of Seiberg-Witten theories with massive
%hypermultiplets,''
Nucl.\ Phys.\ B {\bf 674}, 3 (2003)
[arXiv:hep-th/0212020].
%%CITATION = HEP-TH 0212020;%%
}
%\ManinWJ
\lref\ManinWJ{
Y.~I.~Manin and P.~Zograf,
%``Invertible cohomological field theories and Weil-Petersson volumes,''
arXiv:math.ag/9902051.
%%CITATION = MATH-AG 9902051;%%
}

%\FlumeNC
\lref\FlumeNC{
R.~Flume, R.~Poghossian and H.~Storch,
%``The coefficients of the Seiberg-Witten prepotential as intersection numbers
%(?),''
arXiv:hep-th/0110240;
%%CITATION = HEP-TH 0112211;%%
%\FlumeKB
%``The Seiberg-Witten prepotential and the Euler class of the reduced moduli
%space of instantons,''
Mod.\ Phys.\ Lett.\ A {\bf 17}, 327 (2002)
[arXiv:hep-th/0112211].
%%CITATION = HEP-TH 0110240;%%
}

%\HollowoodDS
\lref\HollowoodDS{
T.~J.~Hollowood,
%``Calculating the prepotential by localization on the moduli space of
%instantons,''
JHEP {\bf 0203}, 038 (2002)
[arXiv:hep-th/0201075];
%%CITATION = HEP-TH 0201075;%%
%``Testing Seiberg-Witten theory to all orders in the instanton expansion,''
Nucl.\ Phys.\ B {\bf 639}, 66 (2002)
[arXiv:hep-th/0202197].
%%CITATION = HEP-TH 0202197;%%
}

%\DijkgraafFC
\lref\DijkgraafFC{
R.~Dijkgraaf and C.~Vafa,
%``Matrix models, topological strings, and supersymmetric gauge theories,''
Nucl.\ Phys.\ B {\bf 644}, 3 (2002)
[arXiv:hep-th/0206255];
%%CITATION = HEP-TH 0206255;%%
%``On geometry and matrix models,''
%%CITATION = HEP-TH 0207106;%%
Nucl.\ Phys.\ B {\bf 644}, 21 (2002)
[arXiv:hep-th/0207106];
%``A perturbative window into non-perturbative physics,''
arXiv:hep-th/0208048.
%%CITATION = HEP-TH 0208048;%%
}

%\CarforaGI
\lref\CarforaGI{
M.~Carfora and A.~Marzuoli,
%``Conformal modes in simplicial quantum gravity and the Weil-Petersson volume
%of moduli space,''
Adv.\ Theor.\ Math.\ Phys.\ {\bf 6}, 357 (2003)
[arXiv:math-ph/0107028].
%%CITATION = MATH-PH 0107028;%%
}

%\MaciociaPH
\lref\MaciociaPH{
A.~Maciocia,
%``Metrics On The Moduli Spaces Of Instantons Over Euclidean Four Space,''
Commun.\ Math.\ Phys.\ {\bf 135}, 467 (1991).
%%CITATION = CMPHA,135,467;%%
}

%\SonnenscheinHV
\lref\SonnenscheinHV{ J.~Sonnenschein, S.~Theisen and
S.~Yankielowicz,
%``On the Relation Between the Holomorphic
%Prepotential and the Quantum Moduli in SUSY Gauge Theories,''
Phys.\ Lett.\ B {\bf 367}, 145 (1996) [arXiv:hep-th/9510129].
%%CITATION = HEP-TH 9510129;%%
T.~Eguchi and S.~K.~Yang,
%``Prepotentials of $N=2$ Supersymmetric Gauge Theories and Soliton Equations,''
Mod.\ Phys.\ Lett.\ A {\bf 11}, 131 (1996)
[arXiv:hep-th/9510183].
%%CITATION = HEP-TH 9510183;%%
E.~D'Hoker, I.~M.~Krichever and D.~H.~Phong,
%``The renormalization group equation in N = 2 supersymmetric gauge theories,''
Nucl.\ Phys.\ B {\bf 494}, 89 (1997)
[arXiv:hep-th/9610156].
%%CITATION = HEP-TH 9610156;%%
}

%\MatoneNF
\lref\MatoneNF{
M.~Matone,
%``Quantum Riemann Surfaces, 2-D Gravity And The Geometrical Origin Of Minimal
%models,''
Mod.\ Phys.\ Lett.\ A {\bf 9}, 2871 (1994)
[arXiv:hep-th/9309096].
%%CITATION = HEP-TH 9309096;%%
}

%\AganagicDB
\lref\AganagicDB{
M.~Aganagic, A.~Klemm, M.~Marino and C.~Vafa,
%``The topological vertex,''
arXiv:hep-th/0305132.
%%CITATION = HEP-TH 0305132;%%
}

%\MarinoCN
\lref\MarinoCN{
M.~Marino and N.~Wyllard,
%``A note on instanton counting for N = 2 gauge theories with classical gauge
%groups,''
arXiv:hep-th/0404125.
%%CITATION = HEP-TH 0404125;%%
}

%\NekrasovVW
\lref\NekrasovVW{
N.~Nekrasov and S.~Shadchin,
%``ABCD of instantons,''
arXiv:hep-th/0404225.
%%CITATION = HEP-TH 0404225;%%
}
%\FucitoUA
\lref\FucitoUA{
F.~Fucito and G.~Travaglini,
%``Instanton calculus and nonperturbative relations in N = 2 supersymmetric
%gauge theories,''
Phys.\ Rev.\ D {\bf 55}, 1099 (1997)
[arXiv:hep-th/9605215].
%%CITATION = HEP-TH 9605215;%%
N.~Dorey, V.~V.~Khoze and M.~P.~Mattis,
%``Multi-instanton check of the relation between the prepotential F and the
%modulus u in N = 2 SUSY Yang-Mills theory,''
Phys.\ Lett.\ B {\bf 390}, 205 (1997)
[arXiv:hep-th/9606199].
%%CITATION = HEP-TH 9606199;%%
}

%\AtiyahRI
\lref\AtiyahRI{
M.~F.~Atiyah, N.~J.~Hitchin, V.~G.~Drinfeld and Y.~I.~Manin,
%``Construction Of Instantons,''
Phys.\ Lett.\ A {\bf 65}, 185 (1978).
%%CITATION = PHLTA,A65,185;%%
}

%\WittenIG
\lref\WittenIG{ E.~Witten,
%``On The Structure Of The Topological Phase Of Two-Dimensional Gravity,''
Nucl.\ Phys.\ B {\bf 340}, 281 (1990).
%%CITATION = NUPHA,B340,281;%%
}

%\DijkgraafDJ
\lref\DijkgraafDJ{
R.~Dijkgraaf, H.~Verlinde and E.~Verlinde,
%``Topological Strings In D < 1,''
Nucl.\ Phys.\ B {\bf 352}, 59 (1991).
%%CITATION = NUPHA,B352,59;%%
}

%\DHokerJM
\lref\DHokerJM{
E.~D'Hoker and D.~H.~Phong,
%``Strong coupling expansions of SU(N) Seiberg-Witten theory,''
Phys.\ Lett.\ B {\bf 397}, 94 (1997)
[arXiv:hep-th/9701055].
%%CITATION = HEP-TH 9701055;%%
K.~Ito, C.~S.~Xiong and S.~K.~Yang,
%``Seiberg-Witten theory as d < 1 topological strings,''
Phys.\ Lett.\ B {\bf 441}, 155 (1998)
[arXiv:hep-th/9807183].
%%CITATION = HEP-TH 9807183;%%
J.~D.~Edelstein and J.~Mas,
%``Strong coupling expansion and Seiberg-Witten-Whitham equations,''
Phys.\ Lett.\ B {\bf 452}, 69 (1999)
[arXiv:hep-th/9901006].
%%CITATION = HEP-TH 9901006;%%
}

\lref\Chen{B.~Chen,
%``Compactification of Moduli Space of Instantons and Its
%Application,''
arXiv:math.gt/0204287.}

\lref\Parker{T.~H.~Parker and J.~G.~Wolfson,
%``Pseudo-holomorphic maps and bubble trees,"
J.\ Geom.\ Anal.\ {\bf 3}, 63 (1993).} \lref\Ekedahl{T.~Ekedahl,
S.~K.~Lando, M.~Shapiro and A.~Vainshtein,
%``Hurwitz
%numbers and intersections on moduli spaces of curves,"
Invent.\ Math.\ {\bf 146}, 297 (2001). S.~K. Lando,
%``Ramified coverings of
%the two-dimensional sphere and intersection theory in spaces of
%meromorphic functions on algebraic curves,"
Russ.\ Math.\ Surv.\ {\bf 57}, 463 (2002).}

\lref\Landoo{S.~K.~Lando and D.~Zvonkine,
%``Counting ramified coverings and
%intersection theory on spaces of rational functions I,"
math.ag/0303218.}

\lref\Kazaryan{M.~E.~Kazaryan and S.~K.~Lando, ``Towards the
intersection theory on Hurwitz spaces," MPI preprint.}

%\TakhtajanVT
\lref\TakhtajanVT{
L.~A.~Takhtajan,
%``Liouville theory: Quantum geometry of Riemann surfaces,''
Mod.\ Phys.\ Lett.\ A {\bf 8}, 3529 (1993)
[arXiv:hep-th/9308125].
%%CITATION = HEP-TH 9308125;%%
} %\KlemmQS
\lref\KlemmQS{ A.~Klemm, W.~Lerche, S.~Yankielowicz and
S.~Theisen,
%``Simple singularities and N=2 supersymmetric Yang-Mills theory,''
Phys.\ Lett.\ B {\bf 344}, 169 (1995) [arXiv:hep-th/9411048].
%%CITATION = HEP-TH 9411048;%%
%}
%\ArgyresXH
%\lref\ArgyresXH{
P.~C.~Argyres and A.~E.~Faraggi,
%``The vacuum structure and spectrum of N=2 supersymmetric SU(n) gauge theory,''
Phys.\ Rev.\ Lett.\  {\bf 74}, 3931 (1995) [arXiv:hep-th/9411057].
%%CITATION = HEP-TH 9411057;%%
} \lref\ScottWolpert{S.~A.~Wolpert, ``Geometry of the
Weil-Petersson completion of the Teichm\"uller space,"
http://www.math.umd.edu/$\sim$saw/preprints/WPcompletion.pdf.}

%references
\Title{}
{\vbox{\centerline{The Liouville Geometry of ${\cal N}=2$ Instantons}
\smallskip
\centerline{and the Moduli of Punctured Spheres}}}
\centerline{Gaetano Bertoldi${}^1$, Stefano Bolognesi${}^2$, Marco
Matone${}^3$, Luca Mazzucato${}^4$ and Yu Nakayama${}^5$}
\smallskip

\centerline{${}^1$ School of Natural Sciences, Institute for
Advanced Study } \centerline{\it Einstein Drive, Princeton, NJ
08540, USA}
\smallskip
\centerline{${}^2$ Scuola Normale Superiore} \centerline{\it P.zza
Dei Cavalieri 7, 56126 Pisa, Italy, and INFN sezione di Pisa,
Italy}
\smallskip
\centerline{${}^3$Dipartimento di Fisica ``G. Galilei'',
Universit\`a di Padova} \centerline{\it Via Marzolo 8, 35131
Padova, Italy, and INFN sezione di Padova, Italy}
\smallskip
\centerline{${}^4$International School for Advanced Studies
(SISSA/ISAS)} \centerline{\it Via Beirut 2 - 4,
 34014 Trieste, Italy, and INFN, sezione di Trieste, Italy}
\smallskip
\centerline{${}^5$Department of Physics, Faculty of Science,
University of Tokyo} \centerline{\it Hongo 7-3-1, Bunkyo-ku, Tokyo
113-0033, Japan}

\noindent We study the instanton contributions of ${\cal N}=2$
supersymmetric gauge theory and propose that the instanton moduli
space is mapped to the moduli space of punctured spheres. Due to
the recursive structure of the boundary in the
Deligne-Knudsen-Mumford stable compactification, this leads to a
new recursion relation for the instanton coefficients, which is
bilinear. Instanton contributions are expressed as integrals on
$\overline{\M}_{0,n}$ in the framework of the Liouville F-models.
This also suggests considering instanton contributions as a kind
of Hurwitz numbers and also provides a prediction on the
asymptotic form of the Gromov-Witten invariants.
 We also interpret this map in terms of the geometric engineering
approach to the gauge theory, namely the topological A-model, as
well as in the noncritical string theory framework. We speculate
on the extension to nontrivial gravitational background and its
relation to the uniformization program. Finally we point out an
intriguing analogy with the self-dual YM equations for the gravitational
version of $SU(2)$ where surprisingly the same Hauptmodule of the
SW solution appears.

%\vskip 0.5cm
%
%
\Date{May 2004}

%
%%%%%%%%%%%%%
%\draftmode
\baselineskip14pt

%
%%%%%%%%%%%%%%%%%%%%%%%%%%%
%
% Table of contents
\centerline{\bf Contents}\nobreak\medskip{\baselineskip=12pt
 \parskip=0pt\catcode`\@=11

\noindent {1.} {Introduction} \leaderfill{2} \par \noindent
\quad{1.1.} {Instantons, moduli of punctured spheres and recursion
relations} \leaderfill{2} \par \noindent \quad{1.2.} {The Stringy
point of View} \leaderfill{4} \par \noindent \quad{1.3.} {Outline
of the Paper} \leaderfill{6} \par \noindent {2.} {Classical
Liouville theory and Weil-Petersson volumes} \leaderfill{7} \par
\noindent \quad{2.1.} {Liouville theory and uniformization of
punctured spheres} \leaderfill{8} \par \noindent \quad{2.2.}
{Deligne-Knudsen-Mumford compactfication} \leaderfill{10} \par
\noindent \quad{2.3.} {Weil-Petersson volume recursion relation}
\leaderfill{11} \par \noindent \quad{2.4.} {The equation for the
Weil-Petersson volume generating function} \leaderfill{13} \par
\noindent \quad{2.5.} {A surprising similarity} \leaderfill{14}
\par \noindent {3.} {The Liouville F-models and the master
equation} \leaderfill{15} \par \noindent \quad{3.1.} {The
Liouville background} \leaderfill{16} \par \noindent \quad{3.2.}
{Intersection theory and the bootstrap} \leaderfill{18} \par
\noindent \quad{3.3.} {The master equation and bilinear relations}
\leaderfill{18} \par \noindent \quad{3.4.} {Pure Liouville quantum
gravity} \leaderfill{19} \par \noindent {4.} {Instanton moduli
space and $\overline \M_{0,n}$} \leaderfill{21} \par \noindent
\quad{4.1.} {Stable compactification and the bubble tree}
\leaderfill{22} \par \noindent \quad{4.2.} {The Hurwitz moduli
space} \leaderfill{23} \par \noindent \quad{4.3.} {The geometry of
Weil-Petersson recursion relations} \leaderfill{25} \par \noindent
{5.} {$\N =2$ gauge theory as Liouville F-models} \leaderfill{26}
\par \noindent \quad{5.1.} {A master equation in $\N=2$ SYM?}
\leaderfill{26} \par \noindent \quad{5.2.} {Relation to ADHM
construction} \leaderfill{28} \par \noindent {6.} {The bilinear
relation} \leaderfill{29} \par \noindent \quad{6.1.} {Inverting
differential equations} \leaderfill{30} \par \noindent \quad{6.2.}
{From trilinear to bilinear} \leaderfill{31} \par \noindent
\quad{6.3.} {The $\N=2$ bilinear relation} \leaderfill{31} \par
\noindent \quad{6.4.} {The Seiberg-Witten solution}
\leaderfill{33} \par \noindent {7.} {Geometric engineering and
noncritical strings} \leaderfill{34} \par \noindent \quad{7.1.}
{Geometric engineering} \leaderfill{34} \par \noindent \quad{7.2.}
{Noncritical string} \leaderfill{37} \par \noindent {8.}
{Discussion} \leaderfill{42} \par \noindent \quad{8.1.}
{Connection with Gromov-Witten theory} \leaderfill{43} \par
\noindent \quad{8.2.} {Graviphoton corrections} \leaderfill{45}
\par \noindent \quad{8.3.} {Higher rank gauge groups}
\leaderfill{49} \par \noindent {9.} {Conclusion} \leaderfill{50}
\par \noindent Appendix {A.} {The restriction phenomenon and the
WP divisor} \leaderfill{51} \par \noindent \quad{\hbox {A.}1.}
{Wolpert's theorem} \leaderfill{51} \par \noindent \quad{\hbox
{A.}2.} {Moving punctures and $D_{WP}^{(n)}$} \leaderfill{52} \par
\noindent References \leaderfill{54} \par

\catcode`\@=12 \bigbreak\bigskip}
%\draftmode

\vfill\eject

%\tenpoint
\newsec{Introduction}

The Seiberg-Witten (SW) solution \SeibergRS\ of the $\N=2$ super
Yang-Mills (SYM) theory is the cornerstone of the today's
nonperturbative understanding of the gauge theories. They realized
that the monodromy property of the coupling constant around the
physical singular points on the moduli space completely determines
the prepotential of the low energy effective theory. SW theory is
characterized by a Legendre duality \MatoneRX, whose precise
structure is determined by the specific $U(1)_{\cal R}$ anomaly
fixing the automorphisms of the fundamental domain for the
$u$-plane \BonelliRY.

There have been along the years many attempts to reconstruct the
complete SW solution from a direct instanton computation (for a
review see \DoreyIK). Originally, the calculation based on the
ADHM construction \AtiyahRI\ has been problematic. Nevertheless,
it has been gradually understood \BellisaiBC\FlumeNC\HollowoodDS\
that the instanton amplitudes are topological objects to which
localization theorems may be applied. In the end, the
all-instanton computation based on the localization technique has
been performed by Nekrasov \NekrasovQD\NekrasovRJ. However, this
topological nature of the instantons and their localization
properties are hard to see just by looking at the instanton
moduli space. The final goal would be to give an
algebraic-geometrical formulation of the instanton moduli space
and its volume form such that one can directly figure out their
localization features. One of the ways in which the instanton
moduli space displays its elegant and deep structure is the
appearance of recursion relations. It has been known for a long
time that the Seiberg-Witten solution obeys some interesting
recursion relations among the coefficients of its instanton
expansion. Different relations are available in the literature,
namely trilinear ones \MatoneRX, extension to higher rank groups \EdelsteinSP, WDVV-like ones
\BonelliQH\MarshakovAE\ItoZR, linear \ChanGJ\ and contact terms
\LosevTP \MarinoGJ. These recursion relations have been attributed to the
underlying topological nature of the instanton moduli space or to
the integrable hierarchies hidden in the SW theory.

\subsec{Instantons, Moduli of Punctured Spheres and Recursion
Relations}

In this paper we study a map between the instanton moduli space of
${\cal N}=2$ SYM for $SU(2)$ gauge group and the moduli space of
the punctured Riemann spheres. This reconstruction of the
instanton moduli space in terms of the moduli space of the
punctured sphere is the main results of this paper. Actually, the
formulation based on the moduli space of the punctured spheres has
an advantage in expressing some algebraic-geometrical feature of
the moduli space of instantons. The natural K\"ahler form on the
moduli space of punctured sphere is known as the Weil-Petersson
(WP) two-form. This WP metric on the punctured spheres not only
yields the natural metric on the moduli space, but also reveals a
remarkable property which is known as the Wolpert restriction
phenomenon. The Wolpert restriction phenomenon guarantees a
localization of integral on the boundary of the moduli spaces for
some particular integrands.

More precisely, we will show some evidence that the moduli space
${\frak M}_n^I$ of the $n$-instanton is mapped to the moduli space
of the sphere with $4n+2$ punctures, namely
$$
{\frak M}_n^I\longrightarrow\overline\M_{0,4n+2}\ ,
$$
by reconstructing the instanton moduli space from SW solution in
terms of the moduli space of punctured spheres. In order to
establish the precise map we will exploit the Liouville
description of the $\overline\M_{0,n}$ spaces. A feature of the WP
volumes is the appearance of a bilinear recursion relation between
them, which is due to the Deligne-Knudsen-Mumford (DKM)
compactification of the moduli space together with the Wolpert
restriction phenomenon. One way to catch this feature of the
moduli space of punctured spheres is to describe it in terms of
Liouville theory. It turns out in fact that the classical
Liouville action is the K\"ahler potential for the WP metric. It
has been found in \MatonePZ\ that this bilinear recursive
structure of the integrals of WP forms on the spaces $\overline
\M_{0,n}$ is preserved if we slightly deform the WP volume forms
$\omega_n$ and evaluate instead of the usual WP volume
$$
\int_{\overline \M_{0,n}}{\omega_n}^{n-3}\ ,
$$
a deformed volume in which we replace the last insertion
with an arbitrary closed two-form
$$
\int_{\overline \M_{0,n}}{\omega_n}^{n-4}\wedge\omega^F \ .
$$
This deformation has been called Liouville F-models. The Liouville
F-models are defined as rational intersection theories on
$\overline{\M}_{0,n}$ and regarded as a certain universality class
of the string theory in a wider sense. This model was originally
advocated to describe the nonperturbative aspects of pure quantum
Liouville gravity in the continuum language. We will see in this
paper that $SU(2)$ SW solution is another example.

The evaluation of such integral defines the expectation value of
the two-form $\omega^F$ in what we denote as the Liouville
background. The benefit of such formulation is that this
expectation values obey a master equation. Our master equation
refines the original formulation of the Liouville F-models in
\MatonePZ, and all the recursive structures of the integral
including its coefficients are now captured by differential
operators $F_n$ which characterize the master equation.

The identification of the coefficient of the $n$-instanton
amplitude with an integral over the moduli space of punctured
spheres described by the Liouville F-models implies that the
instanton moduli space inherits the algebraic-geometrical
properties of the former, in particular its recursive structure,
related to the DKM compactification. Therefore on the basis of
this construction we should expect to find a bilinear recursion
relation among instanton coefficients that shares the common
properties with the one among the WP volumes. Indeed, we find from
the SW solution that the coefficients of the instantons satisfies
the following bilinear relation \eqn\sisisis{ {\F_n\over 2\pi
i}={4n-3\over n}\sum_{k=1}^{n-1}e_{k,n}\F_{k}\F_{n-k}\ . } This
analogy between WP volumes and instantons can be traced back to an
apparent surprising coincidence. On the WP side, a nonlinear ODE,
implied by the bilinear recursion relation for the WP volumes, can
be written as the inverse of a linear differential equation,
satisfied by the generating function of such volumes. On the other
hand, in the ${\cal N}=2$ gauge theory we have a linear
differential equation for the periods, the Picard-Fuchs equation
(PF), whose inverse is a nonlinear ODE and gives the recursion
relation among instanton coefficients, following the other way
around. If we now identify the effective gauge coupling constant
$\tau(a)$ with the `coupling constant' for the generating function
of the WP volumes, we are led to a map between the $\N=2$ SYM
theory and the WP volumes,\foot{A similar map \MatoneWH\ exists
between the $\N=2$ SYM theory and the $\N=1$ SYM theory in the
framework of the Dijikgraaf- Vafa (DV) correspondence \DijkgraafFC
\DijkgraafPP.} the latter being described in terms of the
classical Liouville theory.

In the literature there are well known cases of explicit maps to
$\overline\M_{g,n}$ that simplify considerably the calculations. A
remarkable example is the Hurwitz space. This is the space of
meromorphic functions defining ramified coverings, e.g. of the
sphere. This space admits a compactification $\overline\H_{g,n}$
consisting of stable meromorphic functions \Ekedahl. In particular
the projection $\H_{g,n}\longrightarrow \M_{g,n}$ extends to
\eqn\estenddeAAA{\overline\H_{g,n}\longrightarrow\overline\M_{g,n}
\ . } As we will see, explicit results are simply obtained thanks
to such a map. In particular, calculations simplify considerably
in the genus zero case.

\subsec{The Stringy Point of View} Our construction of the
instanton amplitudes based on the bilinear recursion relation and
Liouville F-models not only provides the reconstruction of the
moduli space of instanton in terms of the punctured spheres, but
also reveals several connections to the stringy setup for the
$\N=2$ SYM theory. In particular, we discuss mostly the connection
to the geometric engineering approach and the noncritical string
approach in this paper.

Let us begin with the instanton amplitudes again. It is now
well-established from the direct instanton calculation that the
the integrands localize in the moduli space \BellisaiBC. This
statement has a natural counterpart in our construction of the
 instanton moduli space in terms of the punctured spheres.
Here the DKM boundary of the moduli space plays an significant
r\^ole in evaluating the integral. Furthermore, the bilienar
recursion relation also suggests the dynamical selection of the
boundary. Namely, we will show that the boundary of the moduli
space which contributes the amplitude consists of the divisor
which separates the number of punctures by multiple of 4 ($+2$),
which nicely fits the naive expectation that the boundary of the
instanton moduli space is given by the collision of two (or more)
`instantons'.

Interestingly enough, the boundary which counts the punctures by
four units can be further reduced to that of one unit. In this
case, the DKM compactification perfectly works and the nature of
the recursion relation is now much like the topological gravity.
We would like to interpret the framework of punctured sphere as
the geometric engineering approach to the $\N=2$ SYM theory. One
can engineer the ${\cal N}=2$ SYM theory by considering the
topological A-model on a certain noncompact Calabi-Yau manifold
and the gauge instanton coefficients are given by the world sheet
instantons wrapping some cycles inside the threefold. Here we
propose instead a worldsheet approach in which the full
topological A-model is considered as a perturbation around the
theory which is obtained in the $a\to\infty$ limit, which
corresponds to the semiclassical limit in the gauge theory ($a$ as
usual denotes the expectation value of the Higgs). The
perturbation is achieved by deforming the world sheet CFT. Since
the gauge theory prepotential in a flat background is given by the
tree level (sphere) free-energy of the A-model, we obtain in this
way the instanton coefficients as integrals on the moduli space of
$n$-punctured spheres. This construction can be easily generalized
to the presence of a nontrivial gravitational background, namely
the graviphoton, whose corrections to the prepotential have
recently raised much attention \NekrasovQD, \FlumeRP.

The direct consequence of our recursion relation and construction
of the instanton amplitudes in terms of the moduli space of
punctured spheres is the prediction for the asymptotic form of the
Gromov-Witten invariants for the local Hirzebruch surface which
yields the $\N=2$ $SU(2)$ SYM in a certain limit. Our bilinear
recursion relation indeed predicts the rescaled version of the
bilinear recursion relation for the asymptotic growth of the
Gromov-Witten invariants. Furthermore, our construction of the
instanton amplitudes in terms of the intersection theory on the
moduli space of punctured spheres states that the asymptotic
growth of the Gromov-Witten invariants is calulable as the
rational intersection theory on $\overline{\M}_{0,n}$.

On the other hand, the quantum Liouville theory (or $c=0$
noncritical string; for a review see \NakayamaVK) is very akin to
the supersymmetric gauge theory in a sense. The supersymmetric
theory depends holomorphically on the parameter, and this property
contributes largely to its solvability and so does the Liouville
theory. Specifically, the method to derive the correlation
functions in the Liouville theory (Goulian-Li \GoulianQR,
Dorn-Otto \DornXN, Zamolodchikov-Zamolodchikov \ZamolodchikovAA)
reminds us of the instanton calculation in the supersymmetric
gauge theory (see, e.g. footnote 36 of \NakayamaVK, see also
\NakayamaEP), where we calculate the amplitude when the
perturbative expression makes sense and then we analytically
continue to the general cases by utilizing the symmetry argument.
Actually, there {\it is} a direct relation between them in the
world sheet $\N=2$ super Liouville theory. In such a theory, the
Liouville superpotential can be derived from the $U(1)$ vortex
condensation, i.e. the instanton effects in the two-dimensional space, from
the parent $\N=2$ $U(1)$ gauged linear sigma model \HoriAX.
Therefore, the dependence of the cosmological constant in
correlation functions is essentially the instanton effect in this
perspective. Since it has been conjectured
\AganagicQJ\DijkgraafFC\ that the bosonic noncritical string
theory is deeply related to the topological twist of the $\N=2$
super Liouville theory,\foot{The point is that the $\N=2$ super
Liouville theory appears in the description of singular Calabi-Yau
spaces and the topological B-model, which is describable in terms
of matrix models, is related to such Calabi-Yau spaces.} the
dependence of the cosmological constant in the bosonic Liouville
theory may have the same origin. In this paper, we push forward
this idea and obtain a more direct connection by using our new
bilinear recursion relation. We rewrite the instanton contribution
to the gauge theory prepotential as the genus expansion of a
certain noncritical string theory, which we propose to call
`instanton string theory'. This theory has a striking resemblance
to the $c=0$ Liouville theory in its structure; actually they are
in the same universality class of the Liouville F-models. The most
intriguing feature is that the amplitude comes only from the
boundary of the moduli space much like in the topological gravity
\WittenHR\DijkgraafDJ. The bilinear recursion relation is nothing
but the string equation in this perspective.

\subsec{Outline of the Paper}

The organization of this paper is as follows. In section 2, we
review the basic facts on the uniformization property and the
moduli space of the punctured spheres. This part consists of the
mathematical foundation of the paper. We will show the crucial relation
between Liouville theory and the WP volumes and we will introduce
the DKM compactification of the moduli space of punctured spheres which,
together with the Wolpert restriction phenomenon, leads to the bilinear recursion
relation of the WP volumes.

In section 3, we introduce the Liouville F-models and the notion
of Liouville background. The former is a certain universality class
of the string theory and comes naturally equipped with the bilinear recursive
structure. We will present a master equation which provides us a
general scheme to treat such bilinear structures in the theory.

In section 4, we discuss the relation between the moduli space
of gauge theory instantons and that of the punctured spheres. We propose
from the algebraic-geometrical perspective
that the former should
be described in terms of the latter. We also discuss some stable
compactifications of moduli spaces and an example of a map to the moduli space
of punctured spheres, namely the map from the space of meromorphic functions on Riemann surfaces,
related to the Hurwitz numbers.

In section 5, we describe the instanton
coefficients as integrals over the moduli space of the punctured spheres.
We introduce a particular Liouville F-model whose master equation predicts the existence
of a bilinear recursion relation, which will be found in section 6 starting from the PF
equations of SW theory.

In section 6, we present our final form of the Liouville F-model
for the $SU(2)$ $\N =2$ SYM. For this purpose, we derive the
bilinear recursion relation hidden in the SW solution. While this
relation is anticipated from our discussion so far, here we derive
it explicitly with precise coefficients. For the existence of the
bilinear recursion relation, we show that the inverse of the PF
potential must be at most quadratic.
 Finally, as a side remark, we point out that if we begin with the bilinear recursion
relation {\it ansatz} with the one-instanton coefficient, we can
rederive even the SW solution itself.

In section 7, we discuss the physical interpretation of this
bilinear relation from the geometric engineering point of view
\KatzFH\ as well as from the noncritical string theory
perspective. In the former approach, by expressing the instanton
amplitudes as integrals on the moduli space of $n$-punctured
spheres, we derive the perturbed CFT expression for the geometric
engineering topological A-model. In the latter approach, we show
that the gauge coupling constant can be written as the second
derivative of a certain noncritical string theory. All these
different approaches are based on the underlying Liouville theory.

In section 8 we propose some speculations and future directions.
We first discuss the possible dualities among various approach to
the SW theory in our view point based on the Liouville geometry.
Then we show the extension to the graviphoton background and the
relation of our bilinear relation to the underlying recursive structure of the
Gromov-Witten invariants.
On the relation to the graviphoton background, we point out an
intriguing analogy with the self-dual YM equations for the gravitational
version of $SU(2)$.
 Finally, we also speculate on
the extension of our results to the higher rank gauge theories.

In section 9 we address some concluding remarks.

In Appendix we report the simple proof of Wolpert's restriction phenomenon
 and the derivation of the Weil-Petersson divisor.

\newsec{Classical Liouville Theory and Weil-Petersson Volumes}

Classical Liouville theory describes the uniformization geometry
which is at the heart of the theory of Riemann surfaces (see e.g.
\MatoneTJ\ for an essential account). In this section we review
the basic properties of the classical Liouville theory including
its r\^{o}le in the description of the geometry of moduli spaces
of Riemann surfaces. As we will see, the Liouville action
evaluated at the classical solution, $S_{cl}$, describes the
metric properties of the moduli spaces just as the Poincar\'e
metric describes the ones of the Riemann surfaces. More precisely,
it turns out that $S_{cl}$ is the K\"ahler potential for the
Weil-Petersson (WP) metric, which, once moduli deformations are
described in terms of holomorphic quadratic differentials, can be
seen as the `natural' metric on moduli spaces.\foot{It turns out
that the WP two-form is also in the same cohomological class of
the Fenchel-Nielsen two-form (see Appendix).} This should be
compared with the r\^{o}le of the Poincar\'e metric whose
logarithm corresponds with its K\"ahler potential (see the
Liouville equation below). Therefore, the Liouville action
describes the geometry of both the Riemann surface, providing its
natural metric given by the equation of motion, and of their
moduli spaces, just coinciding, at its critical point, with the
K\"ahler potential for the WP metric. In this way, roughly
speaking, $S_{cl}$ `transfers' the metric properties of the
Riemann surfaces to their moduli space.\foot{It would be
interesting to investigate whether this important property of the
Liouville action of generating the metric of both spaces may hold
also for other theories.}

\subsec{Liouville Theory and Uniformization of Punctured Spheres}

Here we are mainly interested in the punctured Riemann spheres
\eqn\sfere{\Sigma_{0,n}=\widehat
{\CC}\backslash\{z_1,\ldots,z_n\}\ ,} where $\widehat {\CC}\equiv
{\CC}\cup\{\infty\}$. Since three punctures can be fixed by a
$PSL(2,{\CC})$ transformation, different complex structures may
arise only for $n\geq3$. Let us introduce the moduli space of the
punctured Riemann spheres \eqn\modulisp{ {\cal M}_{0,n}=
\{(z_1,\ldots,z_{n})\in \widehat{\CC}^{n}|z_j\ne z_k\; {\rm for}\;
j\ne k\}/Symm(n)\times PSL(2,{\CC})\ , } where ${Symm}(n)$ acts by
permuting $\{z_1,\ldots,z_n\}$ whereas $PSL(2,{\CC})$ acts as a
linear fractional transformation. We use the latter
transformations to set $z_{n-2}=0$, $z_{n-1}=1$ and $z_{n}=\infty
$, so that we have \eqn\mdls{ {\cal M}_{0,n}\cong
V^{(n)}/{Symm}(n)\ , } where \eqn\star{
 V^{(n)}=\{(z_1,\ldots,z_{n-3})\in
{\CC}^{n-3}|z_j\ne 0,1\ ; z_j\ne z_k\ ,\; {\rm for}\; j\ne k\}\ .
}

A fundamental object in the theory of Riemann surfaces is the
uniformizing mapping \eqn\mapping{ J_\HH: {\HH} \longrightarrow
\Sigma_{0,n}\ ,}
 with ${\HH}=\{w|{\rm Im}\, w>0\}$ the
upper-half plane. The Poincar\'e metric $ds^2={|dw|^2 \over ({\rm
Im}\, w)^2}$ on ${\HH}$, is the metric of constant scalar
curvature $-1$. Since $w= J_\HH^{-1}(z)$, this induces on the
Riemann surface the metric $ds^2=e^{\varphi}|dz|^2$, where
\eqn\metric{ e^{\varphi}={|J_\HH^{-1'}|^2 \over ({\rm Im}\,
J_\HH^{-1})^2}\ . } The fact that the metric has constant curvature
$-1$ is the same of the statement that $\varphi$ satisfies the
Liouville equation \eqn\liouville{\partial_{\bar
z}\partial_z\varphi={e^{\varphi}\over 2}\ .} An important object
is the Liouville stress tensor
 \eqn\bbb{
T(z)=\left\{J_\HH^{-1}(z),z\right\}=\varphi_{zz}-{1\over 2}
\varphi_{z}^2\ ,} where $\{f(z),z\}=f'''/f'-{3\over2}(f'/f'')^2$
is the Schwarzian derivative. In the case of the punctured Riemann
spheres we have \eqn\fucsse{T(z)=\sum_{k=1}^{n-1}\left[{1\over
2(z-z_k)^2}+ {c_k\over z-z_k}\right]\ , } where the {\it accessory
parameters} $c_1,\ldots, c_{n-1}$ are functions on $V^{(n)}$. They
satisfy the two conditions \eqn\condition{ \sum_{j=1}^{n-1}c_j=0\
, \qquad \sum_{j=1}^{n-1} z_jc_j =1-{n\over 2}\ . } Let us write
down the Liouville action \zota\ \eqn\vvv{
 S^{(n)}=\lim_{r\to
0}\left[\int_{\Sigma^r_{0,n}}
\left(\partial_z\varphi\partial_{\bar
z}{\varphi}+e^{\varphi}\right)+ 2\pi (n {\log}
r+2(n-2){\log}|{\log}r|)\right]\ , } where \eqn\ecco{
\Sigma^r_{0,n}=\Sigma_{0,n}\backslash\left(\bigcup_{i=1}^{n-1}
\{z||z-z_i|<r\}\cup\{z||z|>r^{-1}\}\right)\ . } It turns out that
the accessory parameters are strictly related to $S^{(n)}$
evaluated at the classical solution. More precisely, we have the
Polyakov conjecture (see also \CMS\ for a recent discussion)
\eqn\risulta{ c_k=-{1\over2\pi}\partial_{z_k}S_{cl}^{(n)}\ , }
which has been proved in \zota. Furthermore, it turns out that
$S_{cl}$ is the K\"ahler potential for the Weil-Petersson two-form
\zota\ \eqn\wp{ \omega_{WP}^{(n)}= {i\over
2}{\overline\partial}{\partial}S^{(n)}_{cl}=-i\pi\sum_{j,k=1}^{n-3}
 {\partial c_k\over \partial {\bar z_j}}d\bar z_j\wedge d z_k\ ,
} showing that Liouville theory describes the geometry of the
moduli space of Riemann surfaces. As a result, note that even in critical
string theory the classical Liouville action appears in the string
measure on moduli space.

\subsec{ Deligne-Knudsen-Mumford Compactification }

We now consider the DKM stable compactification $\overline
V^{(n)}$ of the moduli space $V^{(n)}$ \DeligneKnudsenMumford. Let
us consider a nontrivial cycle of $\Sigma_{0,n}$. In shrinking it
the complex structure of $\Sigma_{0,n}$ changes and so we move
around $V^{(n)}$. In the limit, when the cycle is completely
shrunk, the degenerate surface does not belong to $V^{(n)}$. A
similar situation arises when one tries to collide two punctures.
In the stable compactification there appears a long tiny neck
separating them with the final configuration corresponding to two
surfaces glued by a node (a double puncture), both having a number
of punctures $\geq3$, so that the Riemann surfaces involved are
always negatively curved. Once the node is removed, we obtain two
Riemann spheres with $k+2$ and $n-k$ punctures, where the value of
$k=1,\ldots,n-3$ depends on the number of encircled punctures.
This sort of exclusion principle, according to which punctures
never collide,\foot{Note that each of the two punctures of the
node belong to different surfaces and in the Poincar\'e metric
their distance is infinity.} is essentially a consequence of the
Gauss-Bonnet theorem.

The DKM boundary of $V^{(n)}$ consists of the moduli spaces
corresponding to such degenerate configurations. Since the
dimension of $V^{(n)}$ is $n-3$, one sees that the boundary has
codimension one. Furthermore, while the number of punctures
increased by two, the total Euler characteristic is unchanged.

A basic property of the DKM compactification is that it has a
clear recursive structure \medskip
$${\eqalign{ &\qquad \overline V^{(k+2)}
\longrightarrow \overline V^{(j+2)} \times \overline V^{(k+2-j)}
\ldots \cr &\nearrow \cr \overline V^{(n)} \longrightarrow
\overline V^{(k+2)} \times \overline V^{(n-k)} & \cr & \searrow
\cr & \qquad \overline V^{(n-k)} \longrightarrow \overline
V^{(j+2)} \times \overline V^{(n-k-j)}\ldots \cr }}
$$
\medskip

\noindent This recursive structure already suggests that for some
suitable forms one may get the localization property
\eqn\localizzazione{\int_{\overline V^{(n)}}\sim
\sum_{k=1}^{n-3}c_k\int_{\overline V^{(k+2)}} \int_{\overline
V^{(n-k)}}\ , } leading to bilinear recursion relation. This
would imply that the generating function for such integrals
satisfy nonlinear differential equations. Therefore from the structure of
the boundary we may get an exact resummation, that is a
`nonpertubative result'. Remarkably, as we will see, in several
cases including $\N=2$ SYM and WP volumes, such equations are
essentially the inverses of linear ones.

\subsec{WP Volumes Recursion Relation}

In order to be formalized, the above description needs a counting
of the number of times a given component $\overline V^{(k+2)}
\times \overline V^{(n-k)}$ appears in the boundary of $\overline
V^{(n)}$. Of course, this is given by the different ways we may
encircle a fixed number of punctures. Thus, we introduce the
divisors $D_1,\ldots,D_{[n/2]-1}$ which are subvarieties of
codimension one, that is they have real dimension $2n-8$. Each
$D_k$ represents, with the above combinatorics, surfaces that
split, upon removal of the node, into two Riemann spheres with
$k+2$ and $n-k$ punctures. In particular, $D_k$ consists of $C(k)$
copies of $\overline V^{(k+2)}\times \overline V^{(n-k)}$ where
\eqn\oddd{C(k)=\pmatrix{n\cr k+1}\ ,} $k=1,\ldots,{(n-3)/2}$, for
$n$ odd. In the case of even $n$ the unique difference is for
$k=n/2-1$, for which we have \eqn\evenn{C(n/2-1)={1\over
2}\pmatrix{n\cr n/2}\ .} It turns out that the image of the
divisors $D_k$'s provides a basis in $H_{2n-8}(\overline{\cal
M}_{0,n},{\RR})$. The DKM boundary simply consists of the union of
the divisors $D_k$'s, that is \eqn\boundary{
\partial\overline V^{(n)}=\bigcup_{j=1}^{[n/2]-1} D_j
\ . } For future purposes it is convenient to extend the range of
the index of $D_j$ by setting \eqn\unomenotre{D_k=D_{n-k-2}\ ,
\qquad k=1,\ldots,n-3\ .} Let us consider the WP volume\foot{Note
that
$${\rm
Vol}_{WP}\left(\overline{\cal M}_{0,n}\right)={1\over n!}{\rm
Vol}_{WP}(\overline V^{(n)})\ .$$} \eqn\volumi{{\rm
Vol}_{WP}\left(\overline {\cal M}_{0,n}\right)={1\over n!(n-3)!}
\int_{\overline V^{(n)}}{\omega_{WP}^{(n)}}^{n-3}\ . } As we will
see, the volumes are rational numbers up to powers of $\pi$, so we
set \eqn\ressccaa{V_n=\pi^{2(3-n)}(n-3)!{\rm Vol}_{WP}(\overline
V^{(n)})\ , } and will consider the rescaled WP two-form
\eqn\rescca{\omega_n={\omega_{WP}^{(n)}\over\pi^2} \ , } as it
will lead to rational cohomology.

In a remarkable paper \Zog, Zograf calculated such volumes
recursively. This construction is simple and elegant. It is
instructive to illustrate the main features leading to his
recursion relation.

\bigskip

\item{(1)} The first step is to note that since the
divisors $D_k$'s provide a basis in $H_{2n-8}(\overline{\cal
M}_{0,n},{\RR})$, the WP two-form $\omega_{n}$ has a Poincar\'e
dual given by a linear combination of the $D_k$'s,\foot{The
derivation of the Poincar\'e dual to $\omega_{WP}$ is reported in
Appendix.} so that $V_n$ reduces to an integral on the boundary of
$\overline V^{(n)}$.

\item{(2)} The recursive structure of the DKM boundary, given by the structure
of the $D_k$'s then implies that $V_n$ is expressed as a sum of
integrals on $\overline V^{(k+2)} \times \overline V^{(n-k)}$.
Thus we start seeing the recursive structure \eqn\seee{
\int_{\overline V^{(n)}}{\omega_{n}}^{n-3}=\sum_{k=1}^{n-3}c_k
\int_{\overline V^{(k+2)}\times \overline V^{(n-k)}}\rho_k\ , }
where $c_k$ are some combinatorial factors.

\item{(3)} The last step is the observation that the
form $\rho_k$ is expressed just in terms of the WP two-forms of
$\overline V^{(k+2)}$ and $\overline V^{(n-k)}$. This is due to
the Wolpert restriction phenomenon \wolpertis, according to which
the restriction of the WP two-form $\omega_{n}$ on each component
$\overline V^{(k+2)}$ of the DKM boundary is in the same
cohomological class of $\omega_{k+2}$. More precisely, we have
\foot{For an excellent updating on the WP metric see
\ScottWolpert.}

\vskip .5cm

\noindent {\bf Theorem} (Wolpert \wolpertis). {\it Let $i$ denote
the natural embedding}

\medskip

 \eqn\embedding{i:
\overline{V}^{(m)}\to\overline{V}^{(m)}\times
* \to \overline{V}^{(m)}\times \overline{V}^{(n-m+2)}
\to \partial \overline{V}^{(n)} \to\overline{V}^{(n)} ,}

\medskip

\noindent {\it for $n>m$, where $*$ is an arbitrary point in
$\overline{V}^{(n-m+2)}\ $. Then}

\medskip

 \eqn\gdtdte{
\left[\omega_{m}\right]= i^*\left[\omega_{n} \right] \ , \qquad
n>m \ . }

\vskip .5cm

\noindent We report in the Appendix a simple proof of this
theorem. This theorem implies that \seee\ becomes \eqn\seeeA{
\int_{\overline V^{(n)}}{\omega_{n}}^{n-3}=\sum_{k=1}^{n-3}\tilde
c_k \int_{\overline V^{(k+2)}}{\omega_{k+2}}^{k-1} \int_{\overline
V^{(n-k)}}{\omega_{n-k}}^{n-k-3}\ . }

\bigskip

\noindent The above is the essential account of Zograf's recursion
relation \Zog\
\eqn\cu{\qquad V_n={1\over 2}\sum_{k=1}^{n-3}{k(n-k-2)\over n-1}
\pmatrix{ n\cr k+1} \pmatrix{n-4\cr k-1}V_{k+2}V_{n-k}\ , } where
$n\geq 4$ and $V_3=1$.

\subsec{The Equation for the WP Volumes Generating Function}

Zograf's recursion relation originated a series of interesting
results in the framework of quantum cohomology \Yuri. The first
observation is that a rescaling of $V_k$ simplifies
Zograf's recursion relation considerably, that is by setting \MatonePZ\
\eqn\rnmdue{
 a_k= {V_k\over (k-1)[(k-3)!]^2}\ ,
} for $k\ge 3$, Eq.\cu\ becomes \eqn\cual{ a_3=1/2\ ,\qquad
a_n={1\over 2}{n(n-2)\over (n-1)(n-3)}
\sum_{k=1}^{n-3}a_{k+2}a_{n-k}\ ,\qquad n\ge 4\ . } This is a
useful step because we can introduce the
generating function of the WP volumes: \eqn\geneterg{
g(x)=\sum_{k=3}^\infty a_k x^{k-1}\ , } and then derive the
associated nonlinear differential equation \MatonePZ\ \eqn\cua{
x(x-g)g''=x{g'}^2+(x-g)g'\ . } The recursion relation \cual\ has
been crucial in formulating a nonperturbative model of Liouville
quantum gravity as a Liouville F-model \MatonePZ\ (see also
\bomama), which we will review in section 3. In particular, this
formulation has been obtained as a deformation of the WP volumes.
Furthermore, it has been argued that the relevant integrations on
the moduli space of higher genus Riemann surfaces can be reduced
to integrations on the moduli space of punctured Riemann spheres
(see also \manin\Givental).

It has been shown by Kaufmann, Manin and Zagier \manin\ that this
nonlinear ODE is essentially the inverse of a linear one. More
precisely, defining $g=x^2\partial_x x^{-1} h$, one has that \cua\
implies \eqn\lalineare{xh''-h'= (xh'-h)h''\ . } Differentiating
\lalineare\ we get \eqn\onan{yy''=xy^3\ , } where $y=h'$. Then,
interchanging the r\^{o}les of $x$ and $y$, one can transform
\onan\ into the Bessel equation
\eqn\bessel{\left(\partial_y^2+{1\over y}\right)x=0\ . }

One known solution of this equation is a modified Bessel function,
which is exactly the inverse of the generating function of the WP
volumes. More precisely, we have \manin\ \eqn\wwrr{x(y)=
-\sqrt{y}J_0'(2\sqrt{y})\ , } where \eqn\bassd{ J_0(z)=
\sum_{n=0}^\infty {(-1)^n\over (n!)^2}\left({z^2\over4}\right)^n\
,} so that \eqn\modified{ y(x)=\sum_{n=3}^{\infty}{V_n \over
(n-2)!(n-3)! } x^{n-2} \quad \Longleftrightarrow \quad
x=\sum_{m=1}^\infty {(-1)^{m-1}\over m!(m-1)!}y^m\ . } The
modified Bessel function is convergent in all the complex plane,
but not its inverse. The first zero of the derivative
$x'(y)=J_0(2\sqrt{y})$ tells us when the inverse stops to
converge. One can calculate the asymptotic form of the WP volumes
in this way.

\subsec{A Surprising Similarity}

There is a surprising similarity between the general structure
involved in the derivation of the WP volumes and the one for the
instantons in $\N=2$ SYM theory for $SU(2)$ gauge group \MatoneRX.
In SW theory one starts from the linear differential equation
satisfied by $a(u)$ and inverts it to a nonlinear differential
equation satisfied by $u=\G(a)=\sum_{k=0}^\infty a^{2-4k}\G_k$.
This in turn implies the recursion relation for the instanton
coefficients $\G_k$ (related to the ones for the prepotential by
$\G_k=2\pi i k\F_k$). In the case of WP volumes, one starts from
the opposite side with respect to SW theory: first directly
evaluates the recursion relation making use of the DKM
compactification and of the Wolpert restriction phenomenon, and
arrives to the nonlinear differential equation \cua\ which should
be compared with the one in $\N=2$ SYM \MatoneRX\
\eqn\quelladea{(1-\G^2)\G''+{a\over4}{\G'}^3=0\ . } What is
crucial is that, like \quelladea, also \cua\ is essentially the
inverse of a linear ODE \bessel.

This stringent analogy strongly suggests the possibility to
reobtain the $\N=2$ SYM results just starting from the point where
Zograf started, that is by directly evaluating the recursion
relation using algebraic-geometrical techniques in instanton
theory. This would be possible in the framework of the Liouville
F-models introduced in \MatonePZ. In this respect we note that,
besides the recursive nature of the DKM compactification and of
the Wolpert restriction phenomenon, two steps leading to
localization phenomena, one of the observations was that the
original recursion relation is actually reduced to the simple form
\cual. Furthermore, a key step was the observation that the
recursive structure obtained by Zograf admits an important
generalization. Namely, in \MatonePZ\ it was observed that the
recursive structure arising in the evaluation of the integrals
$\int_{\overline\M_{0,n}}{\omega_{n}}^{n-3}$ persists even if one
considers the replacement \eqn\hfter{
{\omega_{n}}^{n-3}\longrightarrow
{\omega_{n}}^{n-4}\wedge\omega^F\ , } for suitable closed two-forms
$\omega^F$. This led to the nonperturbative formulation of
Liouville quantum gravity in the continuum \MatonePZ.

\newsec{The Liouville F-models and the Master Equation}
In this section, we consider the Liouville F-models introduced in
\MatonePZ. The Liouville F-models are defined as a certain
universality class of the string theory given by the integration
over the moduli space of the punctured spheres. For example as it
was discussed in \MatonePZ, the $c=0$ noncritical string theory is
in this class. Later we will see that also the SW instanton
contributions are in this category.

The general motivation in \MatonePZ\ to introduce the Liouville
F-models is based on the following observation. A general theory
of string can be represented as a summation on the genus $g$, of
integrals on the moduli space of Riemann surfaces $\overline\M_g$.
Usually such integrals are technically impossible to evaluate.
However, for some theories, these integrals may actually be
performed. In particular, when the theory leads to a localization
in the measure, one expects contributions from the DKM boundary of
$\overline\M_g$, which is composed of the union of copies of the
spaces $\overline\M_{g-1,2}$ and
$\overline\M_{g-k,1}\times\overline\M_{k,1}$, $k=1,\ldots, g-1$.
On the other hand, depending on the structure of the integrand, it
may happen that also each one of the integrals on these lower
dimensional spaces obtains contributions only from the relative
boundary. Iterating this would lead to a final configuration
corresponding to integrals on the moduli space of Riemann surfaces
without handles and with punctures, which is obtained when all the
cycles around the handles are pinched. In this case, the initial
integration on $\overline\M_g$ reduces to a sum of products of
integrals on the moduli space of punctured spheres
$\overline\M_{0,n}$. As a result, after a complete shrinking, the
relevant geometry of the string genus expansion collapses to the
one of $\overline\M_{0,n}$, that is \eqn\roughly{
\{\overline\M_g\}\qquad\Longrightarrow\qquad \{\overline\M_{0,n}\}
\ , } and for such theories one has to evaluate integrals of the
kind \eqn\innntr{ \int_{\overline\M_{0,n}}\rho \ , } for some form
$\rho$. Furthermore, the metric properties of the moduli space
that become the natural measure are described in terms of the WP
two-form, which in turn has the classical Liouville action as
K\"ahler potential. In this respect, we note that the appearance
of the classical Liouville action suggests that for such theories
there is a relation between localization properties of the measure
and the semiclassical approximation (as it may happen in some
noncritical string theories). Thus, one expects that the theory
should be expressed as \eqn\proprionais{ Z_n^F=\int_{\overline
V^{(n)}}{\omega_{n}}^{n-3}e^{-S_F} \ , } where $S_F$ is the
effective action on the moduli, that is the remnant of the initial
quantum action.

\subsec{The Liouville Background}
If one has the effective action $S_F$ which defines
a closed form $\omega^F = e^{-S_F} \omega_n$ up to exact terms,\foot{
Actually, the condition can be weaker in our purpose: the equality
should hold only after the integration wedged by $\omega_n^{n-4}$
 over the moduli space of the punctured spheres.}
a special feature emerges: the amplitudes $Z_n^F$ show a recursive
structure and this distinguishes among others the Liouville
F-models we define in the following.
 In this respect, we mentioned that the
replacement \hfter\ leads to new recursion relations. Therefore,
as observed in \MatonePZ, we can keep the recursion relation
property of the WP volume form not only for $\omega_n^{n-3}$, but
also for other $(2n-6)$-forms. To understand this, note that if we
make the replacement \hfter\ the volumes $V_n$ are replaced by
\eqn\why{\left[\omega_n^{n-4} \wedge \omega^F \right] \cap
\left[\overline{V}^{(n)}\right]=\left[\omega_n\right]^{n-4} \cap
\left[D^F \right]\ ,} where $\cap$ is the topological cup product
and $D^F$ is the Poincar\'e dual to $[\omega^F]$. Expanding the
latter in terms of the homological basis, \why\ becomes
\eqn\abdue{\sum_{k}b_k\left[\omega_n\right]^{n-4}\cap
\left[\overline V^{(k+2)}\times \overline
V^{(n-k)}\right]=\sum_{k}b_k\left[\omega_{k+2}+\omega_{n-k}\right]^{n-4}\cap
\left[\overline V^{(k+2)}\times \overline V^{(n-k)}\right] \ , }
where we used the Wolpert restriction phenomenon we review in the
Appendix. Since the unique contribution comes from the terms in
the binomial expansion having the correct dimension with respect
to $V^{(k+2)}$ and $V^{(n-k)}$, we have
\eqn\abduerree{\sum_{k}b_k\left[\omega_n\right]^{n-4}\cap
\left[\overline V^{(k+2)}\times \overline
V^{(n-k)}\right]=\sum_{k}b_k\pmatrix{n-4\cr k-1} V_{k+2}V_{n-k}\
.} Thus, for suitable $\omega^F$'s one has recursive relations.

The theories we consider are the ones with an effective action
$S_F$ such that \eqn\espresso{Z_n^F=\langle \omega^F\rangle_n \ ,}
where \eqn\aridaiie{ \langle \sigma\rangle_n\equiv {1\over n!}
\int_{\overline V^{(n)}}{\omega_n}^{n-4}\wedge\sigma \ . } This
notation synthesizes some of the peculiar properties enjoyed by
the moduli space of punctured spheres and of its Liouville
geometry. To explain this, first recall that since
\eqn\reccaall{\omega_n={i\over 2\pi^2}\overline\partial\partial
S_{cl}^{(n)}\ , } we see that the notation \aridaiie\ defines the
evaluation of a given two-form in the Liouville background.
Furthermore, we have seen that for suitable $\omega^F$'s, the
substitution \hfter\ preserves the recursive structure of the
integrals. In this respect the notation involves all these
information, in particular in doing the integration one uses the
DKM compactification and the Wolpert restriction phenomenon,
whereas $\omega_n$ provides the Liouville background.

In order to define the divisor $D^F$, we first observe that to
express the theory in terms of integrals on $\overline V^{(n)}$,
we need its recursive structure rather than the specific values of
the WP volumes. Therefore we introduce the normalized divisors
\eqn\abbiamo{\D_k={D_k \over\langle\sigma_k\rangle_n} \ ,} where
$[\sigma_k]$ is the Poincar\'e dual to $D_k$, so that\foot{Note
that ($\langle\omega\rangle_j\equiv \langle\omega_j\rangle_j$)
$$
\langle\sigma_k\rangle_n={(n-4)!(k+2)(n-k)\over(n-k-3)!
(k-1)!}\langle\omega\rangle_{k+2} \langle\omega\rangle_{n-k}\ .
$$
}
\eqn\aduno{
\langle\mu_k\rangle_n=1 \ , }
with $[\mu_k]$ the dual of $\D_k$.

\subsec{Intersection Theory and the Boostrap}

{}From the above discussion, it is clear that we can regard this
formulation as a deformation of the WP volumes. In particular,
recalling that each divisor $D_k$ is composed of copies of
$\overline V_{k+2}\times \overline V_{n-k}$, labelled by two
different subscripts, as will become clear in the following, a
general deformation is conveniently expressed by introducing two
parameters $s$ and $t$. Thus, we introduce the two-form $\eta(s,t)$
whose class is defined by its Poincar\'e dual \eqn\ccddff{
D_{\eta}=\sum_{k=1}^{n-3}e^{(k+2)s}e^{(n-k)t}\langle\omega^F\rangle_{k+2}
\langle\omega^F\rangle_{n-k}\D_k\ .} To understand the nature of
the class $[\eta(s,t)]$, note that since the theories we are
considering share the recursive geometry of $\overline\M_{0,n}$,
the divisor $D_\eta$ includes the physical information on the
`correlators' on the spaces $\overline \M_{0,k+2}\times \overline
\M_{0,n-k}$. Essentially, once the theory is defined, we should
evaluate the (rational) intersection of the divisor associated to
$\omega^F$ with the DKM boundary. Even if we do not know the
explicit expression of such $\omega^F$, we can use the iterative
structure of $\overline\M_{0,n}$ to define $\omega^F$ on
$\overline\M_{0,n}$ in terms of the ones defined on $\overline
\M_{0,k+2}\times \overline \M_{0,n-k}$. It follows that we only
need the first correlator, that is $\langle\omega^F\rangle_3$, as
initial condition, to recover the full tower of $\omega^F$'s. This
means that we compute $\omega^F$ on $\overline\M_{0,n}$, and
therefore the `$n$-point function' $\langle\omega^F\rangle_n$,
starting from the `three-point function'
$\langle\omega^F\rangle_3$. This sort of boostrap is a direct
consequence of the recursive structure of the DKM compactification
and of the Wolpert restriction phenomenon.

\subsec{The Master Equation and Bilinear Relations} As we have
seen so far, the bilinear recursion relation for the WP volume
\cual\ has a natural generalization: \eqn\general{Z_n^F =
\sum_{k=1}^{n-3} F_n(k+2,n-k) Z^F_{k+2} Z^F_{n-k}\ .} Due to the
structure of the DKM boundary, the coefficients $F_n(k+2,n-k)$
defining the general bilinear recursion relation that can be
obtained by evaluating suitable integrations on
$\overline\M_{0,n}$ can be chosen to be symmetric under $k+2\to
n-k$, so that in general \eqn\thefinalsperiamo{
F_n(k+2,n-k)=\sum_j h_j(n)(k+2)^j(n-k)^j\ .} The coefficients
$F_n$ are however redundant, in particular we are interested in
the class of equivalence of recursion relations that differ each
other only by a trivial rescaling of the terms. A natural choice
to select representatives of such equivalence class is to
set\foot{To be precise, this does not uniquely fix $h_0$. However,
this is irrelevant for our purpose.} \eqn\discusione{h_j=0\ ,
\qquad \forall j <0 \ , \qquad {\rm and} \qquad h_0 \neq 0 \ . }
The general Liouville F-models are defined by the generating
function for the $Z_n^F$ where $[\omega^F]$ is given by the master
equation \eqn\master{
[\omega^F]=F_n(\partial_s,\partial_t)[\eta_0]\ ,} where
$\eta_0\equiv\eta(0,0)$. This equation fixes the recursion
relations once the initial condition, that is the three-point
function $\langle\omega^F\rangle_3$, is given. In terms of the
Poincar\'e dual, the master equation reads \eqn\dddx{
D^F=F_n(\partial_s,\partial_t)D_{\eta_0}\ .} Evaluating the master
equation \master\ on the Liouville background
\eqn\masterduea{\langle\omega^F\rangle_n \, = \,
F_n(\partial_s,\partial_t)\langle\eta_0\rangle_n \ , } and by
\aduno\ \eqn\masterdueab{ \langle\omega^F\rangle_n \, = \,
\sum_{k=1}^{n-3}F_n \langle\omega^F\rangle_{k+2}
\langle\omega^F\rangle_{n-k} \ , } which nicely summarizes the
recursive properties of $\overline\M_{0,n}$. In this respect we
note that such a recursion resembles a sort of generalization to
$\overline\M_{0,n}$ of the Riemann bilinear relations. Actually,
both Eq.\masterdueab\ and the Riemann bilinear relations express
the volume form as a linear combination of a the product of two
lower dimensional integrals.

In the previous analysis, we used the recursive properties of the
DKM compactification to introduce the concept of F-models defined
in terms of a suitable set of cohomological classes $[\omega^F]$
that, evaluated on the Liouville background, enjoys the recursive
properties. Besides the DKM compactification, it is just the
Liouville background that, due to the Wolpert restriction
phenomenon, leads to recursion relations.

\subsec{Pure Liouville Quantum Gravity}

The generating functions for the Liouville F-models are \MatonePZ\
\eqn\Fmodels{ {\cal
Z}^{F,\alpha}(x)=x^{-\alpha}\sum_{n=3}^{\infty}x^n\langle\omega^F\rangle_n\
, } which are classified by $\alpha$, $F_n$ and
$\langle\omega^F\rangle_3$. The Liouville F-models include pure
quantum Liouville gravity, which can therefore be formulated in
the continuum in terms of deformation of the WP volumes whose
geometry is described by the classical Liouville theory \MatonePZ.

One of the distinguishing features of pure quantum Liouville gravity is
that, like in the case of WP volumes, we have $h_j=0$, $\forall
j\neq0$, and the master equation takes the simple form
\eqn\masteriii{ [\omega^F]=h_0[\eta_0]\ .} Evaluating this
equation on the Liouville background gives
\medskip \eqn\mastera{\langle\omega^F\rangle_n =
h_0\langle\eta_0\rangle_n = h_0\sum_{k=1}^{n-3}
\langle\omega^F\rangle_{k+2} \langle\omega^F\rangle_{n-k} \ , }
$n\ge4$, where in the case of pure Liouville gravity \MatonePZ\
\eqn\setnumber{h_0={3\over (12-5n)(13-5n)}\ , \qquad
\langle\omega^F\rangle_3=-{1\over2}\ .} Setting ${\cal Z}(t)={\cal
Z}^{F,{12\over5}}(t^5)$, one sees that from the master equation it
follows that the specific heat of pure Liouville quantum gravity
corresponds to the series of the two-form $\omega^F$ evaluated on
the Liouville background \eqn\ppII{{\cal
Z}(t)=t^{-12}\sum_{n=3}^\infty t^{5n} \langle\omega^F\rangle_n \ ,
} which in fact satisfies the Painlev\'e I \eqn\PPPP{ {\cal
Z}^2-{1\over 3}{\cal Z}''=t\ , } with initial conditions ${\cal
Z}(0)={\cal Z}'(0)=0$. Recalling that
\eqn\reccaall{\omega_n={i\over 2\pi^2}\overline\partial\partial
S_{cl}^{(n)}\ , } we see that the formulation in the continuum of
pure Liouville quantum gravity is expressed in terms of the
Liouville action evaluated at the classical solution.

Since this solution is not the standard perturbative solution, we
briefly discuss its nature before concluding this subsection. The
physical consequences of the model have been investigated in
\bomama. In particular, it turns out that the model corresponds to
the quantum Liouville theory with Einstein-Hilbert action having
an imaginary part $\pi/2$. In other words. Eq.\ppII\ corresponds
to introducing a $\Theta$-vacuum structure in the genus expansion
\bomama \eqn\thets{ {\cal F}(t)=
\sum_{g=0}^\infty\int_{Met_g}{\cal D}he^{-S(h)+i{\Theta\over
2\pi}\int_{\Sigma}R\sqrt h}=
\sum_{g=0}^\infty(-1)^{1-g}\int_{Met_g}{\cal D}he^{-S(h)},\quad
\Theta={\pi \over 2}\ ,} where the specific heat is defined as
${\cal Z}(t) =- \F''(t)$. The effect of the $\Theta$-term is to
convert the expansion into a series of alternating signs which is
Borel summable.

An important point is that the specific heat of the model has a
physical behavior. According to standard thermodynamics, if one
defines, following \mm, the `specific heat' as the second
derivative of the free-energy, it should be negative. In \bomama\
it has been shown that the specific heat is negative for all
$t>0$, whereas in the standard choice \mm\ for the boundary
condition in the asymptotic expansion is always positive for
sufficiently large $t$. It seems that this choice is made in order
to avoid an apparently unphysical behaviour such as the
alternating signs of the asymptotic series.\foot{Note that series
with coefficients having alternating signs can be obtained just by
changing the point of the expansion.} However alternating signs in
perturbation theory cannot have a nonpertubative quantum field
theoretical meaning. In other words this `unphysical behaviour' is
only apparent, that is an effect of the perturbation expansion.
What makes sense are the nonperturbative results which are in
complete agreement with basic physical principles. Thus the
results of the model agree with standard thermodynamics and the
theory is Borel summable.

As emphasized in \bomama, the r\^{o}le of $\Theta$-vacua is
suggestive for string theory in general. This aspect is related to
the structure of the moduli space and to unitarity problems. To
understand the relation between unitarity and the structure of
moduli space one should consider that degenerated surfaces
correspond to Feynman diagrams. The r\^{o}le of $\Theta$-vacua
should follow from a Feynman diagram analysis like applied to the
string path-integral at the boundary of moduli spaces. We also
notice that the presence of $\Theta$-vacua should improve the
convergence of the perturbation theory of critical strings. In
other words one should expect that string perturbation theory with
$\Theta$-vacua converges.

\newsec{Instanton Moduli Spaces and $\overline\M_{0,n}$}

Now let us turn to our main theme, the Liouville geometry of the
$\N=2$ instantons and moduli space of punctured spheres. As we
have briefly seen in the introduction, the derivation of the
prepotential in SW theory from the direct instanton calculation
has been steadily developed (for a review see \DoreyIK).
Originally, despite the existence of the exact solution by SW, the
direct instanton calculation based on the ADHM construction
\AtiyahRI\ has been difficult to perform. Nevertheless, it has
been gradually understood \BellisaiBC\FlumeNC\HollowoodDS\ that
the instanton amplitudes are topological objects to which the
localization theorem may be applied. To fully utilize the
technique of the localization, some desingularization of the
instanton moduli space is necessary, and interesting results have
been obtained by Hollowood in \HollowoodDS\ by introducing the
noncommutative geometry (for a more mathematical treatment, see
\NakajimaUH\ and references therein). Finally the all-instanton
solution from the direct instanton calculation based on the
localization technique is presented by Nekrasov
\NekrasovQD\NekrasovRJ, where by using the so-called $\Omega$
background, the enumerative evaluation of the localized integral
becomes possible.\foot{Recently this method has been extended to
other gauge groups in \MarinoCN\NekrasovVW.}

In spite of several proposals, there remain some open questions,
for example the one concerning the compactification. In
particular, even if the above results provide a major step towards
a better understanding of the instanton moduli space of $\N=2$ SYM
theory, it remains to be uncovered whether the instanton moduli
space might admit a more algebraic-geometrical description. The
first step in such a direction has been considered in
\MatoneNT\MatoneWN\FlumeNC. The main idea there was to reconstruct
the instanton moduli space from the known SW solution. In
particular, the strong analogy between Zograf's derivation of the
recursion relation, which leads to a linear differential equation
and the derivation of the instanton recursion relation from the
linear differential equation, suggested a possibility to construct
a space whose structure and volume form directly implies the
instanton recursion relation. However, in such a construction they
have used the original recursion relation, which is a
trilinear one.

There are several results in algebraic-geometry just due to the
choice of a good compactification of a given moduli space. For
example, as we have seen, the DKM compactification, together with
the Wolpert restriction phenomenon, are the two main ingredients
leading to Zograf's recursion relation. On the other hand, we have
seen that a suitable deformation of the WP volume form still
leads to recursion relations. Therefore, the strong similarities
we have seen in section 2.5
between the derivation of WP recursion relation and the one for
the $\N=2$ instantons may actually be a consequence of a
formulation of instanton theory based on the moduli space of
punctured spheres. This would lead to introducing the DKM stable
compactification and therefore to quantum cohomology for $\N=2$
SYM theory, a fact that may not be considered a surprise since
quantum cohomology has several features of gauge theories.

Besides the above-mentioned similarity between WP volumes and $\N=2$
instantons, we also note the theorem \wp\ for the WP two-form
\eqn\wpaga{\omega_{n}= {i\over
2\pi^2}{\overline\partial}{\partial}S^{(n)}_{cl} \ ,} which
relates the K\"ahler potential of the WP metric to the classical
Liouville action. This is reminiscent of the fact that for the
hyper K\"ahler metric of the instanton moduli space, we have
\MaciociaPH\ \eqn\wpinst{ K(\omega_{I}^{(k)}) = -{1\over4}\int
d^4x x^2 {\rm Tr}\,{F_{mn}^{(k)}}^2 \ ,} where
$K(\omega_{I}^{(k)})$ is the hyper K\"ahler potential for the
natural volume of the instanton moduli space which is deduced from
the functional integral, and we evaluate the integral by
substituting the classical $k$ instanton solution just as in \wpaga.

\subsec{Stable Compactification and the Bubble Tree}

The program of using the stable compactification, and therefore quantum cohomology,
for the instantons has been already considered in the
literature (see for example \Chen). The important quantity here is
the moduli space of stable maps. The original proposal of a
similar compactification for moduli spaces of instantons has been
the one by Parker and Wolfson \Parker. Such a compactification is
referred to as the bubble tree compactification.

Another feature indicating the existence of a stable
compactification for the instanton moduli space is the fact
that in this approach `punctures never collide'. Therefore,
 punctures can be considered like fermions, so that there is
a sort of underlying exclusion principle. This similarity can be
explicitly formulated in the geometrical formulation of quantum
Liouville theory \TakhtajanVT\MatoneNF\ and in the framework of
anyon theories \MatoneYR. The latter corresponds to a problem for
particles with configuration space $\overline\M_{0,n}$ whose
dynamics is described by a quantum Hamiltonian which is naturally
given by the Laplacian with respect to the WP metric (thus giving
a self-adjoint operator). In particular, both in Liouville and
anyon theory one can associate a conformal weight to elliptic
points \MatoneNF\MatoneYR\ that in the limit of infinite
ramification, corresponding to a puncture, just gives the value
$1/2$, which is the weight of a fermion. Therefore, in some
respect we can consider punctures behaving as noncommutative
vertices \eqn\vertices{\{\psi(z_i),\psi(z_j)\}\sim \delta(z_i-z_j)
\ . } This would suggest a possible bridge between the stable
compactification and the one considered by Hollowood with the
noncommutative $U(1)$ \HollowoodDS. We also note that in the ADHM
construction it should be possible, in principle, to find a
suitable embedding of the matrices in $\overline\M_{0,n}$ such
that the degenerated configurations be naturally compactified \`a
la DKM.

The emergence of the fermion here has also an interesting feature in
connection with the recent developments around Nekrasov's solution and
the geometric engineering approach to the $\N=2$ SYM theory. On one hand,
in \NekrasovAF, the $\tau$-conjecture has been proposed, which states
that the full-genus (graviphoton corrected) partition function is described
by the quantum dynamics of chiral fermions on the SW curve. On the other
hand, in \AganagicQJ, it has been discussed that in order to formulate
topological vertex in terms of the mirror B-model, which in the end computes
relevant amplitudes for SW solution, it is useful to introduce chiral fermion
which describes noncompact B-branes.

\subsec{The Hurwitz Moduli Space}

Besides instantons, there is another important theory which maps
to the moduli space of Riemann surfaces, including the one of
punctured spheres. This is the space of meromorphic functions of
degree $n$ on genus $g$ Riemann surfaces defining the degree $n$
ramified coverings of the sphere. In the case in which the poles
are simple and the critical values of the function sum to 0, this
space, denoted by $\H_{g,n}$, is a smooth complex orbifold,
fibered over $\M_{g,n}$, whose fiber is naturally defined by
associating to each function the corresponding Riemann surface.
This space admits a compactification $\overline\H_{g,n}$
consisting of stable meromorphic functions \Ekedahl.
 The basic fact is that the projection
$\H_{g,n}\longrightarrow \M_{g,n}$ extends to
\eqn\estendde{\overline\H_{g,n}\longrightarrow\overline\M_{g,n} \
. } A feature of such a projection is that we do not have a vector
bundle since the dimension of the fiber may vary by varying the
point on the base. One is then interested in the fiberwise
projectivization $P\overline\H_{g,n}$. As such, this space has a
natural two-form given by the first Chern class of the
tautological sheaf \eqn\tautto{\psi_{g,n}=c_1({\cal O}(1))\in
H^2(P\overline\H_{g,n}) \ .} A consequence, which is of interest
for our purpose, is that the space $P\overline\H_{0,n}$ turns out to be
fibred on $\overline\M_{0,n}$, whose fiber is the projective space
$PE$ where \eqn\whitney{ E=\oplus_{k=1}^n L_k^\vee \ , } is the
Whitney sum of the tangent lines to the curve at the punctures.
This means that the cohomological algebra is generated by
$\psi\equiv\psi_{0,n}$ subject to the relation \Landoo\
\eqn\rellxx{\psi^n+ \sum_{k=1}^n\psi^{k-1}c_{k}(E)=0 \ . } It
follows that considering the natural projection \eqn\ppkrtion{
\pi: P\overline\H_{0,n}\; \longrightarrow \; \overline\M_{0,n} \ ,
} we can express any $\alpha\in H^{2d}(P\overline\H_{0,n})$ as
\eqn\nnaaisssg{
\alpha=\pi^*(\eta_d)+\pi^*(\eta_{d-1})\psi+\pi^*(\eta_{d-2})\psi^2+\ldots
\ , } $\eta_k\in H^*(\overline\M_{0,n})$ . Then, since $\pi_*
\psi^s=c_{s-n+1}(-E)$, it follows that the degree of $\alpha$ can
be evaluated in terms of integrals on $\overline\M_{0,n}$
\Kazaryan
 \eqn\osicjj{
{\rm deg}\, \alpha=\int_{P\overline\H_{0,n}}
{\pi^*(\eta_d)+\pi^*(\eta_{d-1})\psi+\pi^*(\eta_{d-2})\psi^2+\ldots\over
1-\psi}=\int_{\overline\M_{0,n}} {\eta_d+\eta_{d-1} +\ldots\over
c(E)} \ ,} whose evaluation turns out to be considerable
simplified. The above result provides an explicit interesting
example that integrals on complicated spaces actually simplify
just due to a natural map to $\overline\M_{0,n}$. This further
supports our program of expressing instanton contributions in
terms of integrals on $\overline\M_{0,n}$. In this respect, what
is of interest is not just the dimensional reduction of the
moduli, as it may also happen that there are convenient and higher
dimensional parametrizations leading to a simplification, rather
we see that $\overline\M_{0,n}$ provides a sort of basic space
where the complicated integrations simplify considerably.

\subsec{The Geometry of WP Recursion Relation}

If one looks for a formalism that expresses the $\N=2$ instanton
contributions as integrals on the moduli space, in such a way that
it respects the bilinear recursive structure of the DKM
compactification, one should first understand how the DKM geometry
may reflect in the expected bilinear recursion relation. In order
to check the existence of such a bilinear recursion relation, we
note that a characteristic feature of the classical Liouville
theory is that the recursion relation for the WP volumes has
particular properties. Let us write down Eq.\cual\ once again
\eqn\cualagain{a_3=1/2\ ,\qquad a_n={1\over 2}{n(n-2)\over
(n-1)(n-3)} \sum_{k=1}^{n-3}a_{k+2}a_{n-k}\ ,} for $n\ge4$. The
bilinear nature of such a recursion relation is a consequence of
the fact that the boundary of the DKM compactification is the
union of the products of {\it two} moduli spaces of lower order.
This seems difficult to reproduce in $\N=2$ SYM theories.
Actually, it was shown in \MatoneWN\ that a possible analogy was
to construct a moduli space for instantons whose boundary has
components which also contain the product of {\it three}
subspaces. To discuss this point in some detail we should first
better understand the geometrical reason underlying the structure
of the recursion relation \cualagain. Recalling that it was
obtained as the DKM boundary contribution to the integral, we have
\eqn\dalbounbdary{ {\rm dim}\,(\overline\M_{0,n})={\rm dim}\,
(\overline\M_{0,k+2})+{\rm dim}\,(\overline\M_{0,n-k})+1\ , }
$k=1,\ldots,n-3$, which is satisfied because ${\rm
dim}\,(\overline\M_{0,n})=n-3$. It follows that the existence of a
formulation in terms of the Liouville F-models, such as expressing
the instanton contributions as integrals on the moduli space of
punctured spheres, would require that the resulting recursion
relation reflects the main properties of the one of WP volumes.
Thus, besides being bilinear, it should have other characteristic
features which follow from \dalbounbdary. In particular the range
of the recursion relation should be given by $k=1,\ldots,n-3$.
Furthermore as suggested by \dalbounbdary\ the indices of the
instanton contributions involved in such a, still hypothetical,
bilinear recursion relation should be $n-k$ and $k+2$. However,
there is a minor subtle point with such an identification. Namely,
note that the above data may change by a simple shift of the
indices by a global constant and by a shifting of $n$. More
precisely, if we define $a_{k}=b_{k+m}$ and rewrite Eq.\cualagain\
for $n+m$, we obtain \eqn\cualagainb{ b_{3+m}=1/2\ ,\qquad
b_n={1\over 2}{(n-m)(n-m-2)\over(n-m-1)(n-m-3)}
\sum_{k=1}^{n-m-3}b_{k+m+2}b_{n-k}\ ,} for $n\ge 4+m$. This shows
that even if it looks different from \cualagain, in fact they have
the same content. This suggests that in the general bilinear
recursion relation \eqn\generale{ d_n=\sum_{k=1}^p c_k
d_{k+q}d_{n-k}\ ,\qquad n\geq r\ , } one introduces the two
quantities that remain invariant under the above transformations,
i.e. \eqn\invariante{\A=p+(k+q)+(n-k)=n+p+q\ ,\qquad \B=r-q-1\ . }
By \cualagain, in the case of the WP volumes we find
\eqn\si{\A_{WP}=2n-1\ ,\qquad \B_{WP}=1\ .}

\newsec{$\N=2$ Gauge Theory as Liouville F-models}

In this section we give an explicit construction of instanton
amplitudes for $\N=2$ $SU(2)$ SYM theory on the basis of the
Liouville F-models. We first guess the correct dimensionality of
the integrand to reproduce the bilinear recursive structure of the
amplitudes. Then we discuss the dynamical selection of the
boundaries which should be related to the localization of the
instanton effects in our formalism. These observations finally
lead to our master equation of the $\N=2$ SYM whose precise form
will be determined in the next section by deriving the bilinear
recursion relation.

\subsec{A Master Equation in $\N=2$ SYM?}

We have seen that there is a lot of evidence for the existence of
a formulation of instanton contributions in terms of integrals on
the moduli space of punctured spheres. We should expect that the
instanton moduli space can be mapped to $\overline\M_{0,n}$ along
the lines of what happens in the case of the Hurwitz space. Since
the number of parameters for a single instanton does not
correspond with the one dimensional complex coordinate of a
puncture, we expect that the map should be to the product of
moduli spaces of punctured spheres of a particular kind. On
dimensional grounds, one should expect that matching of the
parameters should associate $4$-punctures to each instanton, so we
would at first consider the space $\overline\M_{0,4n}$ whose
dimension is $4n-3$. However, this identification leads to
problems with dimensional matching to get the bilinear recursion
relation. Actually, this identification will give \eqn\willf{
\overline\M_{0,4n}\; \longrightarrow \;
\overline\M_{0,4n-k}\times\overline\M_{0,k+2} \ . } Note that
whereas the left hand side refers to a number of punctures which
is a multiple of 4, in the right hand side never appear pairs of
moduli spaces both having a number of punctures which is multiple
of $4$. Therefore, with this choice it would be cumbersome to make
the identification of, say, the instanton contribution $\F_n$ to
the SW prepotential, with an integral over $\overline\M_{0,4n}$,
and simultaneously satisfying a bilinear recursion relation.
However, note that in the case $\overline\M_{0,4n+2}$ we would
have the pattern \eqn\willfggg{ \overline\M_{0,4n+2}\;
\longrightarrow \; \overline\M_{0,4n-k+2}\times\overline\M_{0,k+2}
\ , } so that when $k$ is a multiple of 4 we may identify $\F_n$
with an integral on $\overline\M_{0,4n+2}$ and consistently having
the bilinear recursion relation.

We then set \eqn\poniamoN{
\F_n=\sum_{k=1}^{n-1}a_k\int_{\D_{4k+2}}\sigma_k^{4k-1} \ , }
where $\sigma_k$ is some two-form on $\D_{4k+2}$ to be determined.
In order to understand the nature of such a form we consider the
natural embedding \eqn\dddt{ i: \overline{V}^{(4k+2)}\to
\overline{V}^{(4k+2)}\times * \to \overline{V}^{(4k+2)}\times
\overline{V}^{(4n-4k+2)} \to \partial \overline{V}^{(4n+2)} \to
\overline{V}^{(4n+2)}, \qquad n>k\ , } where $*$ is an arbitrary
point in $\overline{V}^{(4n-4k+2)}$. On the other hand according
to Wolpert theorem we have \eqn\aasappa{
\left[\omega_{4k+2}\right]= i^*\left[\omega_{4n+2}\right], \qquad
n>k\ .} Using \poniamoN\dddt\ and \aasappa\ , we can express
$\F_n$ in the form \eqn\esprimiamooo{ \F_n\sim \int_{\overline
V^{(4n+2)}}\omega_{4n+2}^{\;\; 4n-2}\wedge\omega^F \ , } where
$[\omega^F]$ is the dual of a linear combination of the divisors
$\D_{4k+2}$. The above investigation suggests that

\medskip

\noindent {\it The instanton contributions can be expressed as
integrals on the moduli space of punctured Riemann spheres leading
to a bilinear recursion relation.}

\medskip

\noindent This allows us to write $\F_n$ in terms of the bilinear
recursion of the F-models. Before doing this, we recall that in
deriving the master equation we observed that there is a canonical
way to select a bilinear recursion relation from the ones which
differ by a trivial rescaling of the terms. Therefore, rather than
$\F_n$ we will express a rescaled version $\bar \F_n$. Note that
this is irrelevant as a global rescaling does not change the
structure of the recursion relations. We saw that the divisor
$D^F$ receives contributions only from the moduli spaces of
Riemann spheres with $4k+2$ punctures, that is $F_{N}(k+2,(N-k))$,
$N=4n+2$, vanishes unless $k$ is multiple of 4. Equivalently, we
can consider for $D_\eta$ the expansion \eqn\ccddffdd{
D_{\eta}=\sum_{k=1}^{n-1}e^{(4k+2)s}e^{(N-4k)t}\langle\omega^F\rangle_{4k+2}
\langle\omega^F\rangle_{N-4k}\D_{4k}\ .} From the master equation,
in the case of $\N=2$ SYM we should then have
\medskip \eqn\masterennedueA{\bar
\F_n=\langle\omega^F\rangle_{N} \, = \,
F_N(\partial_s,\partial_t)\langle\eta_0\rangle_{N} \ ,} which
leads to \eqn\masterennedueB{\langle\omega^F\rangle_{N} \, = \,
\sum_{k=1}^{n-1}F_{N}(4k+2,(N-4k)) \langle\omega^F\rangle_{4k+2}
\langle\omega^F\rangle_{N-4k} \ , } $n\geq2$. In the next section
we will show that this relation holds and we will fix both
$F_{N}(4k+2,(N-4k))$ and the initial condition
$\langle\omega^F\rangle_6$.

\subsec{Relation to ADHM Construction} We would like to briefly
comment on the relation of this proposal to the usual treatment of
the integral over moduli spaces of instantons based on the ADHM
construction. Explicit instanton calculations show that there
appear considerable simplifications in performing integrals over
the whole moduli space. This indicates that the original
parametrization might not be the most convenient choice in
specific cases. The $n$-th instanton contribution to the
prepotential for $SU(2)$ can be obtained from the integration over
the ADHM instanton moduli space ${\frak M}_n^I$, whose metric (and
hence the volume form $d^{8n}\mu$) is given by \wpinst, as
\eqn\Instanton{ \F_n = {1\over {\rm Vol}({\RR}^4)}\int_{{\frak
M}_n^I} d^{8n}\mu e^{-S_{eff}}\ ,} where the effective action
$S_{eff}$ is given by the integration over the fermionic
coordinates. In this computation, four out of the total $8n$ real
coordinates of the moduli space are identified with the center of
the instantons. Dividing by the volume of spacetime corresponds to
a regularization of the infrared divergence related to the
integration over the center. The net result is an integration over
the $8n-4$ coordinates left. On the other hand, an interesting
result is that this same integration can be rewritten as an
integration of the $(4n-3)$-power of a two-form \FlumeNC\
\eqn\Instantong{ \F_n \simeq \int_{\widehat{{\frak M}}^I_n}
(d\rho)^{4n-3} \ . } Here the moduli space $\widehat{{\frak
M}}^I_n$ is obtained from the usual ${\frak M}^I_n$ by first
dividing by the center of the instanton as in \Instanton\ and then
by integrating over the complex scale of the ADHM moduli. The
closed two-form $d\rho$ can be formally considered as the Euler
class associated with the original ADHM moduli space ${\frak
M}_{n}^I$ regarded as a $U(1)$ bundle (see \FlumeNC\ for details).
>From a dimensional point of view note that both \Instanton\ and
\Instantong\ have been performed after fixing the center of the
instanton coordinates, but the latter gets rid of an additional
complex coordinate related to a rescaling of the ADHM moduli. Next
we note that it is very useful introducing noncommutative $U(1)$
instantons when regularizing divergences \HollowoodDS, with the
effect of smoothing out the singularities of the moduli space. In
this case it is convenient to keep the four dimensional explicit
integration of the center of the instantons. Therefore it would be
natural to consider a moduli space of real dimension $8n-2$, where
we simply rescale the moduli. This is just the dimension of
$\overline\M_{4n+2}$ we selected before.

 Let us also note that the DKM boundary of the
moduli space of punctured spheres is deeply related to the
negatively curved nature of the punctured spheres (with more than
2 punctures). Stability guarantees that spheres with two punctures
do not appear. We also note that in the hyperbolic metric the
distance between two punctures is infinite. This is in particular
the case of the two punctures that remain on removal of the node.
However, while the two punctures are infinitely separated in the
Poincar\'e metric, they are at zero distance in the Euclidean
metric. This is better understood if one considers the upper-half
plane: the distance between a point in the upper-half plane and
the boundary is divergent as $1/y$ whereas with the Euclidean
metric we just have $y$. The crucial remark is that the product of
the two distances is a constant. On the other hand, from the
Euclidean point of view, as we are approaching the boundary of the
upper-half space, i.e. a puncture, we are considering an infrared
problem, whereas in the hyperbolic metric we are always
considering a long-distance, i.e. infrared, geometry. This dual
picture is just a sort of UV/IR duality. Actually, this indicates
a physical meaning of the DKM compactification, which may lead to
a direct connection with the $U(1)$ noncommutative compactification
by Hollowood \HollowoodDS. We also note that the infrared and
ultraviolet regularization properties of negatively curved
manifolds have been observed in \CallanEM. It is worth also
mentioning that such UV/IR regularization is strictly related to
the distortion theorems for univalent functions (see the second
reference in \MatoneRX).

\newsec{The Bilinear Relation}

Until now we have collected evidence that the instanton
contributions in $\N=2$ SYM for $SU(2)$ gauge group can be
formulated as a deformation of the Liouville theory leading to
integrals over the moduli spaces of punctured spheres.
Nevertheless, apparently there are several reasons that seem to
prevent such a construction. The main obstacle is that the known
recursion relation for the instantons is a trilinear one
\MatoneRX. Let us briefly recall how it was obtained. Consider the
relation between the $u$-modulus and the prepotential \MatoneRX\
\eqn\mato{ u=\pi i\left({\cal F}-a^2{\partial {\cal
F}\over\partial a^2}\right)\ . } This can be derived either by
instanton analysis, \FucitoUA\ therefore without explicitly
calculating the $\F_k$ or $\G_k$, or by means of the
superconformal anomaly \HowePW. Taking the derivative of Eq.\mato\
with respect to $u$, one sees that the periods $a_D(u)$ and $a(u)$
satisfy a second-order differential equation without the first
derivative term, called the PF equations and whose precise
potential is given from the SW curve: \eqn\pf{
\left[\partial_u^2+{1\over4(u^2-1)}\right]a(u)=0\ . } If we invert
\pf\ for $u=\G(a)$ we find the nonlinear equation \eqn\quellade{
(1-\G^2)\G''+{a\over4}{\G'}^3=0. } Expanding in power series
$\G(a)=\sum_{k=0}^\infty \G_k a^{2-4k}$, one finds the trilinear
recursion relation
$$
{\cal G}_{n+1}=
$$
\eqn\trili{ {1\over 2(n+1)^2} \left[(2n-1)(4n-1){\cal
G}_n+\sum_{k=0}^{n-1}c_{k,n}{\cal G}_{n-k}{\cal G}_{k+1}
-2\sum_{j=0}^{n-1}\sum_{k=0}^{j+1}d_{j,k,n}{\cal G}_{n-j} {\cal
G}_{j+1-k}{\cal G}_{k}\right]\ , }
where $n\geq 0$, ${\cal G}_0=1/2$
and
$$c_{k,n}=2k(n-k-1)+n-1
\qquad d_{j,k,n}= [2(n-j)-1][2n-3j-1+2k(j-k+1)]\ .
$$
In this section, we overcome this difficulty and derive the
bilinear recursion relation. As a result we determine
$F_N(4k+2,(N-4k))$ and the initial condition $\langle
\omega^F\rangle_6$ for the master equation of the $\N=2$ SYM
theory.

\subsec{Inverting Differential Equations}

The trilinear recursion relation \trili\ is unsatisfactory for our
purpose because it does not respect the essential bilinear
recursive property of the DKM compactification with which our
master equation naturally is equipped. Therefore it is natural to
ask a question when we can obtain the bilinear recursion relation
starting from the general potential $V(x)$ in the PF equation. In
this subsection and the next, we derive the necessary conditions
to obtain a bilinear recursion relation in this setup.

Since the following construction is general, we will use $\psi(x)$
rather than $a(u)$. Let us consider the second-order differential
equation 
\eqn\second{ \left[\partial^2_x +V(x)\right]\psi(x)=0\ ,
} and set $x=\G(\psi)$. Since $\partial_x={\G'}^{-1}\partial_\psi$
and
$\partial_x^2=-{\G'}^{-3}\G''\partial_\psi+{\G'}^{-2}\partial^2_\psi$,
where here $'\equiv\partial_\psi$, it follows that equation for
$\G(\psi)$ satisfies the differential equation

\eqn\tri{V^{-1}\G''=\psi {\G'}^3\ . }

Now suppose we want to expand $\G(\psi)$ in power series and gain
a recursion relation for the coefficients of this
expansion from \tri. Clearly the relation we will obtain would be trilinear
at least, due to the appearance of the third power of $\G'$.
Fortunately enough, it turns out that for a certain class of functions $V(x)$,
there is a nice way to simplify this nonlinear equation in order
to obtain a bilinear equation. This holds only in the case where
$V^{-1}(x)$ is at most quadratic in its argument.

\subsec{From Trilinear to Bilinear}

Introduce a function $\H(\psi)$ such that $\H'=\G(\psi)$ and define the auxiliary function
$f$ through the following relation \eqn\effe{ V^{-1}+f\H''=0\ , }
so that a factor $\H''$ drops and \tri\ becomes
\eqn\cucco{f\H'''+\psi{\H''}^2=0\ . } Differentiating \effe\ we
have by \cucco\

\eqn\bucchio{
f'\H''-\psi{\H''}^2+\partial_\psi V^{-1}=0\ . } On the other hand,
since $V$ is a function of $\G=\H'$, we have $\partial_\psi
V^{-1}=\H''\partial_{\H'}V^{-1}$, so that \bucchio\ becomes
\eqn\bucchiene{ f'-\psi\H''+\partial_{\H'}V^{-1}=0\ , } which we
can easily integrate to obtain the following expression for the
auxiliary function \eqn\auxi{ f=\psi\H'-\H-\int^\psi
d\tilde\psi\partial_{\H'}V^{-1}\ . } Plugging this solution into
the defining equation \effe\ for $f$, we obtain our final equation
\eqn\bili{ \left(\psi\H'-\H-\int^\psi d\tilde\psi
\partial_{\H'}V^{-1}\right)\H''+ V^{-1}=0\ . } We then have

\medskip

\noindent {\bf Proposition.} {\it By} \bili\ {\it it follows that
if $V^{-1}(x)$ is a polynomial at most quadratic in $x$; then the
recursion relation for the coefficients of the power expansion of
$x=\G(\psi)$, with $\psi(x)$ solution of} \second{\it, is
bilinear}.

\medskip

\subsec{The ${\cal N}=2$ Bilinear Relation}

We now readily apply this procedure to the PF equation \pf.
By introducing $\H'(a)=\G(a)$ and following the above steps we
find the following nonlinear equation
 \eqn\bisiacca{
1-{\H'}^2={1\over4}\H''(a\H'-9\H)\ . } Note the while \quellade\
was trilinear, this last \bisiacca\ is just bilinear. Plugging the
asymptotic expansion \eqn\essppa{
\H(a)=\sum_{k=0}^\infty{\G_k\over3-4k}a^{3-4k}\ , } into
\bisiacca\ we eventually find the bilinear recursion relation for
$\hat\G_n\equiv\G_n/(4n-3)$ \eqn\recca{ \hat\G_n=
\sum_{k=1}^{n-1}g_{k,n}\hat\G_{k}\hat\G_{n-k}\ , } where $n\geq2$,
$\hat\G_1={1\over2^2}$ and \eqn\valoregnk{
g_{k,n}={(2n-15)k(n-k)\over n^2}+3\ . } This shows in ${\cal N}=2$
SYM we have the master equation \masterennedueA, where we identify
$\bar\F_n=\hat\G_n$ and \eqn\sssiii{ F_N=3+{2n-15\over
16n^2}(\partial_s-2)(\partial_t-2) \ , \qquad
\langle\omega^F\rangle_6={1\over 2^2} \ . }

By equation \mato\ we obtain the relation between the coefficients of the
prepotential $\F(a)=\sum_k a^{2-4k}\F_k$ and $\G_k$, namely $\G_k=2\pi ik\F_k$, $k\geq1$. Then we
obtain the bilinear recursion relation for the instanton expansion
of the prepotential as \eqn\prepu{
{\F_n\over 2\pi i}={4n-3\over
n}\sum_{k=1}^{n-1}e_{k,n}\F_{k}\F_{n-k}\ , } where \eqn\askl{
e_{k,n}={k(n-k)\over (4k-3)[4(n-k)-3]}g_{k,n}\ . }
This completes our construction of the instanton moduli space in terms
of the moduli space of punctured spheres and the consequent derivation of the
master equation for $\N=2$ SYM theory proposed in section 5.
 Also note that
we have the same invariants introduced in section 4.3:
\eqn\uiuilsi{
\A_{WP}=\A_{\N=2}\ , } \eqn\uiuilsib{
\B_{WP}=\B_{\N=2}\ . }

As a final remark, we note that the auxiliary function $\H'=\G$ can be actually integrated
by the PF equation \pf\ to
\eqn\paraculatina{ \H={u\over9}a+{4\over9}(u^2-1)a'\ . } Observe
that $\H$ has the same monodromy of the period $a$. This suggests the
possibility to introduce its dual \eqn\paraculatinaa{
\H_D={u\over9}a_D+{4\over9}(u^2-1)a_D'\ , } satisfying the
equation $\partial_{a_D}\H_D=u$ which is the dual of $\H'=u$. Also note that by
$\partial_{a_D}\H_D=\partial_a\H$ we have \eqn\wwwop{ {\partial
\H_D\over\partial\H}=\tau\ . } Since $\H$ naturally appears in
deriving the bilinear relation, it should be further considered in
the framework of SW theory. Its physical interpretation, which may
appear in the strong coupling region, considered in \DHokerJM, needs to be uncovered.

\subsec{The SW solution}

Before going to the stringy interpretation of our result, we would
like to point out an interesting possibility. Our discussion of
the derivation of the bilinear recursion relation is based on the
known SW solution. However, if we turn the logic around and first
assume the existence of the bilinear recursion relation {\it
without} specifying its coefficients, we can {\it derive} the SW
solution. In other words, if we know the instanton amplitude is
given by a certain (unknown) Liouville F-model beforehand, we can
reproduce the SW solution, and hense specify the actual Liouville
F-model. Our assumption is
\medskip

\noindent {\it The instanton contributions can be expressed as
integrals on the moduli space of punctured Riemann spheres leading
to a bilinear recursion relation.}

\medskip

\noindent We will also make use of the one-instanton contribution,
which is $\G_1=1/2^2$, and of the relation between the $u$-modulus
and the prepotential \MatoneRX\ \eqn\mato{ u=\pi i\left({\cal
F}-a^2{\partial {\cal F}\over\partial a^2}\right)\ . } This relation can be
derived either by instanton analysis, that is proving
$\G_k=2\pi i k\F_k$ directly, \FucitoUA \DoreyZJ\ therefore
without explicitly calculating the $\F_k$ or $\G_k$, or by means
of the superconformal anomaly \HowePW.

{}From the relation \mato, we know that $a(u)$ satisfies the PF
equation with an unknown potential $V(u)$. However, according to
the Proposition, $V^{-1}$ should have the form \eqn\generico{
V^{-1}(u)=Au^2+Bu+C\ .} Recalling that $u=\G(a)=\H'(a)$, we have
by \bili\ \eqn\ggeenn{ \H''[a\H'-(1+2A)\H-Ba]+A{\H'}^2+B\H'+C=0\ .
} Putting the asymptotic expansion \eqn\essppa{
\H(a)=\sum_{k=0}^\infty{\G_k\over 3-4k}a^{3-4k}\ , } with
$\G_0=1/2$, which follows by the asymptotic expansion of $\F(a)$
and Eq.\mato, we obtain \eqn\azero{\sum_{j,k\geq0}\G_j
\G_ka^{4-4(j+k)}\left[4{(2k-1)(2j+A-1)\over
3-4j}+A\right]+B\sum_{j\geq0}\G_j (4j-1)a^{2-4j}+C=0\ . } Now
observe that each term $a^{2-4k}$ is multiplied only by a singular
$\G_k$, so that \eqn\unooon{B=0\ .} Next, the term $a^2$ gives
\eqn\duuud{ A=4 \ .} The coefficient $C$ is determined by
requiring that $\G_1=1/2^2$, which gives \eqn\treeet{C=-4 \ .}
Therefore we have obtained the PF equation for $\N=2$ SYM for
$SU(2)$ gauge group \eqn\pf{
\left[\partial_u^2+{1\over4(u^2-1)}\right]a(u)=0\ . } We have then
shown that the conjectured relation between $\N=2$ SYM and the
Liouville theory leads us to an assumption that the bilinear
recursion relation for $\N=2$ exists, and this assumption together
with the `initial condition', which in the case of $\N=2$ is given
by the values of $\G_0$ and $\G_1$, is sufficient to completely
fix the whole solution.

\newsec{Geometric Engineering and Noncritical
String Theory}

In this section we show that the map from the instanton
moduli space to the moduli space of punctured Riemann surfaces
have a counterpart in
terms of the geometric engineering approach and the noncritical
string theory. These approaches shed new light on the physical origins
of this map.
The similarity of the derivation suggests that our
recursion relation strictly reflects the duality among the different descriptions.

\subsec{Geometric Engineering}

We have seen that we can map the problem of calculating instanton
contributions to integrals on the moduli space of the
punctured spheres. We now discuss how this may be interpreted
in terms of the geometric engineering formulation. Consider the
bilinear recursion relation as in \prepu, which has the same
structure we encounter in the Liouville F-models.
In particular, a nice feature of the
matching of the dimensions in \dalbounbdary\ is that it holds
for a generic decomposition of the moduli space of a $(r+2)$-punctured sphere,
due to its DKM compactification origin. In particular, we can now
identify the $n$-instanton coefficient with an integral over
the moduli space of an $n$-punctured sphere.
Let us consider the Liouville F-models and set $Z_n^F = \F_{n-2}$,
then the recursion relation of $Z_n^F$ \masterdueab\ is translated
as \eqn\Zdef{\F_n = Z_{n+2}^F = \sum_{k=1}^{n-1}
F_{n+2}(k+2,n+2-k)\F_k\F_{n-k}\ .} Now if we set \eqn\setF{{1
\over 2\pi i}F_{n+2}(k+2,n+2-k) = {4n-3 \over n} e_{k,n} \ ,} we
obtain the desired recursion relation \prepu. Note that the
divisor introduced here consists of all the possible number of
punctures, as opposed to that of the previous section, where
the divisor or its coefficients $F(n,k)$ vanish unless $k$ is
multiple of $4$. In the reduction here, we no longer need to
restrict our boundaries to the multiple of $4$, which we believe
is a more compact formulation.\foot{In order to obtain the
canonical form introduced in section 3, we need to rescale $\F_n$
 and use $\hat{\G}_n$ instead. Setting $Z_n^F = \hat{\G}_{n-2}$,
we obtain $F_{n+2}(k+2,n+2-k) = g_{k,n}$.}
Though we have lost the perspective as the integration over the
instanton moduli space, we have a new interpretation of this
integration over the moduli space of the $n$-punctured spheres as
the perturbative calculation of the free-energy in the topological
string theory. Furthermore it yields a
clue to extend our formulation to the case in which the gauge theory
is coupled to a gravitational
background (see
e.g. \IqbalIX\EguchiSJ\ for a comparison of the geometric
engineering expression for the graviphoton corrected prepotential
with Nekrasov's formula). The argument goes as follows.

First, let us recall the geometric engineering approach \KatzFH\
to obtain the Seiberg-Witten prepotential. In this approach, we
consider a topological A-model on a certain local Calabi-Yau space
and the free-energy of this theory yields the whole prepotential
of the $\N=2$ SYM theory. In our $SU(2)$ case, we take the local
Hirzebruch surface (${\PP}^1$ bundle over ${\PP}^1$). Local here
means that we consider the canonical line bundle over the
Hirzebruch surface to construct a noncompact Calabi-Yau threefold.
The proper identification of the parameters between the gauge
theory and the geometry is
 \eqn\geomlimit{e^{-T_B} =
\left({\beta\Lambda \over 2}\right)^4\ ,\qquad T_F =2a\beta\
,\qquad g_s = \beta \hbar \ , } where $T_B$ is the volume of the
base ${\PP}^1$, $T_F$ is that of the fiber and the parameter
$\beta$ is introduced to obtain the four dimensional field theory
limit as $\beta\to0$. This setup can be obtained also by
considering M-theory on the same Calabi-Yau threefold times a
circle of radius $\beta/2\pi$, obtaining a five dimensional gauge
theory with eight supercharges \GopakumarJQ. The four dimensional
field theory limit $\beta\to0$ here corresponds to shrinking the
circle to zero size. Furthermore, the string coupling $\hbar$ in
the A-model is identified with the four dimensional self-dual
graviphoton field strength $F_+$ \AntoniadisZE \BershadskyCX\ and,
after taking the field theory limit, this allows us to identify
the higher genus topological string amplitudes with the
gravitational corrections to the prepotential ${\cal F}(a,\hbar)$
of the ${\cal N}=2$ four dimensional gauge theory
\NekrasovAF\EguchiSJ.\foot{The difference among the various kind of
Hirzebruch surfaces is irrelevant in this limit.} Therefore, if we
restrict ourselves to the usual gauge theory on flat background,
what we need to evaluate is just the genus zero free-energy of the
topological CFT. Particularly the $n$-th gauge theory instanton
contribution comes from worldsheet instantons that wrap $n$ times
the base ${\PP}^1$.

To connect this fact with our expression of the prepotential,
which is described by the integration over the moduli space of the
$n$-punctured spheres, we will consider the A-model from a world
sheet point of view. Consider the period $a$ of the gauge theory and
take the $a\to \infty$ limit of the
above setup, which corresponds to the semiclassical limit in the
field theory, thus obtaining an A-model CFT whose action
we denote as $S_{\infty}$. Then for finite $a$ the world sheet
action can be written as a perturbed CFT with respect to the
original $S_{\infty}$ \eqn\psps{S_{a} = S_{\infty} + {1 \over a^4}
\int d^2z O(z)\ , }
for a certain operator $O(z)$.
 The $1/a^4$ expansion of the SW prepotential
${\cal F}(a)$ will be given by the free-energy of this latter CFT,
which we can evaluate as a series in the perturbation.\foot{ This
is only true if $1/a^4$ is a special coordinate of the topological
CFT. Though it is a difficult problem to verify this, our
formulation suggests that this is indeed true. We also expect that
$O(z)$ is formally BRST exact so that it contributes to the
amplitude only through the contact terms, as we can see from our
recursion relation. The whole construction seems more transparent
if we use the mirror symmetry and move to the B-model, but in the
abstract CFT language used here, there is no essential
difference.} Usually, the A-model free-energy is computed by the
world sheet instanton summation (Gromov-Witten invariants), but
here we would like to regard the A-model just as an abstract
topological CFT instead.\foot{ We postpone the connection with the
Gromov-Witten invariants to the last section.} Now suppose we want
to evaluate the free-energy of this CFT: the perturbative
insertion of the vertex operator ${1 \over a^4} \int d^2z O(z)$
will give us an extra moduli (puncture) and the integration over
its position. Thus this nicely reproduces our expression of the
prepotential as the integration over the moduli space of the
$n$-punctured spheres. Therefore, on the basis of our Liouville F-model analysis,
we expect that

\eqn\gcf{{\cal F}_n \sim \int_{\overline{\cal M}_{0,n+2}}
{\omega_{0,n+2}}^{n-2} \wedge \omega^I \sim\left\langle \left(\int
d^2z O(z) \right)^{n} \right\rangle_{S_\infty} \ .}

The merit of this interpretation is that we know how to
incorporate the graviphoton correction in the A-model side. It is
essentially given by the higher genus free-energy as explained in
\geomlimit\ \AntoniadisZE \BershadskyCX. This provides a clue to
extend our formulation to the gravitational background: we simply
consider the same CFT on higher genus Riemann surfaces. It would
be very interesting to complete this program and we leave it as a
future study. It should be noted, however, that though we expect
that the graviphoton corrected prepotential is given by the
integration over the moduli space of the punctured higher genus
Riemann surfaces, the integrand needs not a priori coincide with
\gcf.

\subsec{Noncritical String}

Finally, we present a noncritical string interpretation of the bilinear
recursion relation. As we have stated in the introduction, the
noncritical string theory and the supersymmetric gauge theory have
many common features. There we have discussed the relation to the
world sheet $U(1)$ instanton. However, when we consider the
space-time theory, things become more interesting. We have learned
that the instanton part of the $d=4$, $\N=2$ SYM coupling constant
$\tau = {1\over 2\pi i}\sum_k \tau_k a^{-4k}$ obeys by \prepu\ the
bilinear recursion relation \eqn\recc{ \tau_n =
 \sum_{k=1}^{n-1} e_{k,n}'
\tau_{k}\tau_{n-k}\ , } where \eqn\newtau{e'_{k,n} =
{(4n-3)(4n-2)(4n-1)\over
n[4(n-k)-1][4(n-k)-2](4k-1)(4k-2)}e_{k,n}\ .} The reason why we consider
$\tau$ instead of $\F$ will be clarified later by the strong
analogy to the $c=0$ Liouville theory where we mainly study the
specific heat, but here we only note that it is dimensionless. We
would like to understand this relation in terms of the string
genus expansion.

It was proposed in
\bomama\ that this kind of recursion
relation can be rewritten as the perturbative noncritical string
form (there it was done for the Painlev\'e I --- the $c=0$
noncritical string theory). Rewriting in this form enabled us to
understand the origin of reduction of the path integration over
the Liouville field into the known cohomological objects of the
moduli space of Riemann surface (see also the recent argument of
Zamolodchikov \ZamolodchikovYB\
for a possible explanation of this reduction mechanism).

Here, rewriting the SYM recursion relation in terms of a string perturbation
theory touches some interesting issues. First, we might find a
confining string theory behind SYM, which we name as instanton string theory.
 Second, we may have a better
understanding on the negative expansion of the cosmological
constant in the Liouville theory.\foot{It is impossible to
understand this in terms of the world sheet instanton theory. The
negative dependence of $\mu$ cannot be understood from the usual
instanton expansion. This is also true in the usual Goulian-Li
approach of the Liouville theory in the higher genus.} It should
also be noted that the proposal that the noncritical string theory
is deeply connected with the $4d$ (supersymmetric) gauge theory
can be found in Polyakov \PolyakovTJ\ before the discovery of
AdS/CFT. He considered a certain noncritical string theory as a
dual description of the YM theory. The usual difficulty is that
the noncritical string theory has a $c=1$ barrier, but he proposed
that in the higher dimension, the Liouville dimension becomes the
warped geometry instead. In our opinion, this proposal has a nice
realization in the $\N=2$ super Liouville theory: thanks to the
discovery of the duality in the $\N=2$ Liouville theory
conjectured in \AhnSX\ and proved in \NakayamaVK\ (see also
\HoriAX \HoriCD\ for a dual picture and a relation to the SW
theory) the Liouville potential has a dual realization as the
warped geometry.\foot{One minor difference, however, is that the
warped coordinate is only the Liouville direction.} The duality
between the B-model on conifold and the $c=1$ noncritical string
theory \GhoshalWM\ is another example of this phenomenon. The
partition function of the conifold describes the universal
nonperturbative effect of the $\N=1$ supersymmetric gauge
theories.

In our $\N=2$ SYM theory, the genus expansion is given by
$\left({\Lambda}\over{a}\right)^{4g}$ in the SYM side. On the
noncritical string side, we expect $g_s^{2g}\mu_r^{-2g}$, where
$\mu_r$ is the renormalized cosmological constant. Thus the
correspondence is \eqn\cores{\eqalign{\Lambda^4=\Lambda_0^4
e^{-4\left({{8\pi^2}\over{g^2}}-i\theta\right)/b}&\sim g_s^{2}=
e^{2\langle \phi\rangle}\ , \cr a^4 &\sim \mu_r^2\ ,} } where $b=
2N_c-N_f=4$ in our particular $SU(2)$ case. Looking at this
expression, we make two observations. On the one hand, the
dependence of the $\theta$ parameter in the SYM side reminds us of
the stringy $\Theta$ parameter \thets\ proposed by
Bonelli-Marchetti-Matone
 \MatonePZ. Secondly, the correspondence between $a$ and
$\mu_r$ may have the geometrical origin. In the $\N=2$ Liouville -
CY correspondence, $\mu_r$ plays the r\^{o}le of the deformation
of the complex moduli parameter, and it should be related to the
moduli of the $4d$ theory --- in this case, of course, the $\N=2$
SYM moduli $a$.

Furthermore, it is interesting to see the connection between the
KPZ scaling law of the noncritical string theory and the
(fractional) instanton contribution of the supersymmetric gauge
theory to the physical quantities. In the KPZ scaling law, the
path integration over the Liouville zero mode leads us to the
genus scaling law of the cosmological constant which we identify
with the SYM moduli. On the other hand, in the $\N=2$ pure SYM
theory, nonperturbative corrections to the prepotential come only
from the instanton and not from fractional instantons in contrast
to the $\N=1$ gauge theories. These structures in the gauge
theories are also determined from symmetry arguments as is the
case with the Liouville theory. This
coincidence should have the same origin as the emergence of the
$\N=2$ Liouville theory as the world sheet theory of the
corresponding gauge theories. In this perspective, the fact that
the amplitudes involving fractional instantons, which is typical
in $\N=1$ gauge theories, usually demand an analytic continuation
corresponds to the fractional power of the cosmological constant
in the Liouville theory, which also demands an analytic
continuation procedure to evaluate correlators.

Now, in order to obtain the noncritical string expression for the
instanton string theory, we will repeat the argument given in \bomama\
 in our context. We consider moduli spaces of higher
genus punctured Riemann surfaces and define the expectation value
in the corresponding Liouville background as\foot{The omitted
suffix to $\sigma$ is the same of the one appearing as suffix of
the bracket. We will use this notation throughout this
subsection.} \eqn\BMMff{ \langle \sigma \rangle_{g,n} \equiv
{{1}\over{(3g-3+n)!}} \int_{\overline{\cal
M}_{g,n}}{\omega_{g,n}}^{3g-2+n} \wedge \sigma\ ,} where we follow
the notations introduced in section 3. This Liouville background
terminology makes sense here because $S_{cl}$ is the potential for
the WP metric even for the punctured higher genus Riemann surfaces
(see \zota). We define the genus expansion of the $\N=2$ effective
coupling constant as\foot{Since prepotential is naturally
connected to the free-energy, we use its second derivative, that
is, the coupling constant here to compare the result with \bomama.
The emergence of $\overline{\cal M}_{g,2}$ is naturally understood
in this context. This moduli space is related to the topological
nature of the recursion relation.} \eqn\BMMf{ \tau_g=\langle
\omega_I\rangle_{g,2}\ . } Let us introduce the basis of divisors
$D_0 = \overline {\cal M}_{g-1,4}$,
 $D_k=\overline {\cal M}_{k,2}\times \overline {\cal M}_{g-k,2}$,
$k=1,\ldots, g-1$. Furthermore it is convenient to rescale the
divisors by the normalized WP volumes as we have done in section 3
so that ($\langle \omega \rangle_{i,j}\equiv \langle \omega_{i,j}
\rangle_{i,j}$)

\eqn\resdiv{ \D_0 = {D_0 \over \langle \omega \rangle_{g-1,4}} \ ,
\qquad \D_k = {D_k \over \langle \omega \rangle_{k,2}} \ .} We now
define a divisor $D_I$, which we call `$\N=2$ divisor', as the
$(6g-4)$-cycle \eqn\NDiv{ D_I= c_0^{(g)}\langle \omega_I
\rangle_{k,4} \D_0 + \sum_{k=1}^{g-1}c_k^{(g)} \langle
\omega_I \rangle_{k,2}\langle \omega_I \rangle_{g-k,2} \D_k
\ ,} where the coefficients $c_k^{(g)}$ will be given later. We
identify $\left[\omega_I\right]$ as the Poincar\'e dual to
$D_I$, i.e.
$\left[\omega_I\right]=c_1\left([D_I]\right)$ where,
as usual, $[D]$ denotes the line bundle associated to a given
divisor $D$ and $c_1$ is the first Chern class. Our rescaling of
the divisor is based on the same spirit in section 3.

We now fix the $c_k^{(g)}$'s by requiring that $\tau_g$'s defined
in \BMMf\ satisfy the recursion relation \recc. Two facts are
crucial to obtain recursion relation: first, in evaluating the
relevant integrals will appear only the components of the boundary
$\partial {\overline{\cal M}}_{g,2}$ of the form $\overline{\cal
M}_{g-k,i}\times \overline{\cal M}_{k,j}$ with $i=j=2$ and
$\overline{\cal M}_{g-1,4}$, second, $\omega_{g,2}$ satisfies the
restriction phenomenon mentioned above. In particular, considering
the natural embedding
\eqn\shutup{ i: {\overline{\cal M}_{k,2}}\to
{\overline{\cal M}_{k,2}}\times * \to {\overline{\cal
M}_{k,2}}\times {\overline{\cal M}_{g-k,2}} \to
\partial {\overline{\cal M}_{g,2}}\to {\overline{\cal M}_{g,2}}\ , }
$g>k$, where $*$ is an arbitrary point on $\overline
{\M}_{g-k,2}$, one has by the Wolpert theorem \wolpertis
\eqn\hotrats{ \left[\omega_{k,2}\right]=i^*
\left[\omega_{g,2}\right]\ , } and similarly for $\omega_{g-1,4}$.
By Poincar\'e duality one obtains
\eqn\three{ \langle
\omega_I\rangle_{g,2} = {1\over (3g-1)!}
\left[\omega_{g,2}\right]^{3g-2}\cap \left[D_I\right]\ , }
so that by \shutup\ and \hotrats\ and setting
\eqn\sette{c_0^{(g+1)}=0\ , } we have \eqn\six{\tau_{g}= {1\over
(3g-1)!} \sum_{k=1}^{g-1}{c}_k^{(g)} \langle \omega_I
\rangle_{k,2}\langle \omega_I \rangle_{g-k,2} \left[
\omega_{k,2}+\omega_{g-k,2}\right]^{3g-2} \cap \D_k \ .}
 The only contribution
in the RHS comes from (recall $\D_k$ is proportional to
$\overline {\cal M}_{k,2}\times \overline {\cal M}_{g-k,2}$)
\eqn\avviene{ {\omega_{g-k,2}}^{3(g-1-k)+2}\wedge
{\omega_{k,2}}^{3k-1}\ , } therefore we have \eqn\seven{
\tau_{g}= {1\over 3g-1}\sum_{k=1}^{g-1}c_k^{(g)}\tau_{k,2}
\tau_{g-k,2}\ , }

The recursion relation \seven\ coincides with \recc\ if we set
\eqn\eight{c_k^{(g)} =(3g-1)e_{k,{g}}'\ ,} for $k>0$. Notice that all the
coefficients $c^{(g)}_k$ are rational numbers so that $D_I$
defines a rational homology class and the above computations can
be interpreted in the sense of rational intersection theory.

Substituting \seven\ into $\tau(a) = {1\over 2\pi i}\sum_{g}\tau_g
a^{-4g}$, we have asymptotically
\eqn\iuhdw{ \tau(a)\sim {\rm
perturbative \ part}+{1\over 2\pi i}\sum_{g=1}^\infty
a^{-4g} \langle \omega_I\rangle_{g,2}\ .}

So far we have obtained the instanton coefficients as formal
integration of
cohomological objects over the moduli space of the Riemann
surface. If we want to identify this as a string perturbation
theory, we should address the question of finding a worldsheet
action for the noncritical string theory. In this respect
note that this structure arises as Liouville F-models. In our context,
the difference is only the coefficient of the divisor. Therefore
we expect that the instanton string theory for $\N=2$ SYM is given
by a certain noncritical string theory though we cannot give a
precise action at this moment. This perspective will eventually
connect the $\N=2$ SYM theory with a noncritical string theory as
was proposed by Polyakov. Furthermore, in this view, we can naturally
understand the origin of the bilinear recursion relation as the string
equation of the instanton string theory.

For a preliminary study of this connection, we propose a possible
world sheet reduction mechanism of the recursion relation as has
been done in the case of $c=0$ Liouville theory. For this purpose
we give a conjectural argument that could relate our {\it ansatz}
\BMMf\ to the path-integral approach to our instanton string
theory. In this line of thought a key step is the
Duistermaat-Heckman (DH) theorem \Duistermaat\ which roughly
speaking corresponds to the following
statement.\foot{Duistermaat-Heckman theorem is also essential in
the instanton calculation. See e.g. \DoreyIK\ and references
therein. Thus if we study the connection with our formalism here,
it will lead to some information on the reduction mechanism in the
noncritical string theory.} Let $X$ be a $2n$-dimensional
symplectic manifold with symplectic form $\omega$ and $H$ a
Hamiltonian on $X$. Then integrals such as \eqn\rere{ {1\over
n!}\int_X \omega^n e^{-H}\ ,} only depend on the behavior of the
integrand near the critical points of the flow of the Hamiltonian
vector field. The point is that in a path-integral approach one
expects that the contribution at genus $g$ to the two-puncture
correlation function of instanton string theory which we identify
as the coupling constant (because it is the second derivative of
the free-energy) is given by\foot{In this subsection we use the
notation $\langle{\,}\rangle^{\rm CFT}$ when referring to CFT
correlators as opposed to the usual $\langle\,\rangle$ notation
for the Liouville background.} \eqn\qkudgw{ \langle{{\cal
O}_0}^2\rangle_g^{ {\rm CFT}}=
\langle\omega\, e^{-H}\rangle_{g,2} \equiv {1\over (3g-1)!}
\int_{\overline {\cal M}_{g,2}}{\omega_{g,2}}^{3g-1}e^{-H}\ , }
where $H$ is an `effective action' arising from the integration in
the path-integral at fixed moduli in $\overline {\cal M}_{g,2}$.
The two-form $\omega_{g,2}$ is symplectic on $\overline {\cal
M}_{g,2}$, regular in the interior and extending as a current to
the boundary, therefore, regarded as a map from $T^* \overline
{\cal M}_{g,2}$ to $T \overline {\cal M}_{g,2}$,
${\omega_{g,2}}^{-1}$ has zeroes only on $\partial\overline {\cal
M}_{g,2}$. Furthermore, since $\omega_{g,2}$ is a K\"ahler form,
the Hamiltonian vector field is given by ${\omega_{g,2}}^{-1}dH$
so that the flow of the Hamiltonian vector field has critical
points at $\partial\overline {\cal M}_{g,2}$. Let us assume that
DH applies to the integral \qkudgw\ and furthermore it gets
contributions only from the critical points in the component of
$\partial\overline {\cal M}_{g,2}$ whose factor contain an even
number of punctures. Then one expects
$$
\langle{{\cal O}_0}^2\rangle_{g}^{ {\rm CFT}}\sim \alpha_{g}
\langle \omega\, e^{-H}\rangle_{g-1,4} +\sum_{k=1}^g\beta_{g,k}
\langle \omega\, e^{-H}\rangle_{k,2}\langle
\omega\, e^{-H}\rangle_{g-k,2} $$ \eqn\joegarage{ =
\alpha_{g}\langle{{\cal O}_0}^4\rangle_{g-1}^{ {\rm CFT}}+
\sum_{k=1}^{g-1}\beta_{g,k} \langle{{\cal O}_0}^2\rangle_{k}^{ {\rm
CFT}} \langle{{\cal O}_0}^2\rangle_{g-k}^{ {\rm CFT}}\ ,} where
$\alpha_{g}$, $\beta_{g,k}$ are possibly $a$-dependent
coefficients. The difference from the $c=0$ Liouville theory
studied in \bomama\ is that $\beta$ has a $k$ dependence. Let us
introduce the cohomology classes $\left[\sigma_{g,k}\right]\in
H^{2}\left({\overline{\cal M}_{g,2}}\right)$, $k=0,\ldots, g-1$,
Poincar\'e dual of $\D_k$ defined in \resdiv.
 Introducing the normalized WP volumes $\langle \omega_{g,n}\rangle_{g,n}$
and using the fact that $\langle{{\cal O}_0}^4\rangle^{\rm CFT}$ is related
to the second derivative of $\langle{{\cal O}_0}^2\rangle^{ {\rm CFT}}$, one
obtains for the asymptotic behavior of the correlation functions
\eqn\joegaragetwo{\langle{{\cal O}_0}^2\rangle_{g}^{ {\rm CFT}}\sim
\tilde{\alpha}_{g} \langle{{\cal
O}_0}^2\rangle_{g-1}^{ {\rm CFT}} \langle\sigma_{g,0}\rangle_{g,2}
 + \sum_{k=1}^{g-1}\beta_{g,k} {\langle{\cal
O}_0^2\rangle_{k}^{ {\rm CFT}}}{\langle{{\cal
O}_0}^2\rangle_{g-k}^{ {\rm CFT}}} \langle\sigma_{g,k}\rangle_{g,2} \ ,
}
where $\tilde{\alpha}_g = \alpha_g{25(g-1)^2-1\over 4}$.

 Since the asymptotic expression of $\langle{{\cal
O}_0}^2\rangle_{g}^{ {\rm CFT}}$ evaluated at $a=1$ is equal to ${1\over 2\pi
i}\tau_{g}$, by setting $a=1$ in \joegarage\ we derive \eqn\gdteb{
\tau_{g}\sim \langle \omega^{(g,2)}\rangle_{g,2}\ ,
} with $\omega^{(g,2)}$ a two-form given by \eqn\enormousthree{
\omega^{(g,2)}=\tilde{\alpha}_{g}{{\langle \omega^{(g-1,2)}\rangle_{g-1,2}}}
\sigma_{g,0}-\sum_{k=1}^{g-1}\beta_{g,k} {\langle\omega^{(k,2)}\rangle_{k,2} }
{\langle\omega^{(g-k,2)}\rangle_{g-k,2} }\sigma_{g,k} \ , }
exhibiting the same structure of the two-form $\omega_I$
introduced in the {\it ansatz} \BMMf\ with the coefficients
$c_k^{(g)}$ given by \eight.

Particularly, note that $\tilde{\alpha}_{g} = 0$. This should have an
interesting physical origin. Moreover this construction clearly
explains the natural emergence of the normalization by the
WP volume factor in the recursion relation from the localization
technique.

\newsec{Discussion}

So far, we have obtained the three different expressions for the
prepotential (or the effective gauge coupling constant), starting
from our bilinear relation: the integration over the moduli space
of the $(4n+2)$-punctured spheres, the integration over the moduli
space of the $n$-punctured spheres, and the genus expansion
(`instanton string theory'). These expressions are physically
interpreted as the direct instanton calculation, the topological
A-model expansion, and the noncritical string theory respectively.
We emphasize that the three different approaches to the SW theory
is deeply connected via our bilinear recursion relation.

Here we would like to discuss the nature of the duality among
these expressions from our viewpoint based on the Liouville geometry.
The duality
between the topological A-model approach (geometric engineering)
and Nekrasov's approach (instanton counting) has been recently
studied in \IqbalIX \EguchiSJ\ microscopically. {}From our point
of view, the essential reason is that the integration over the
both moduli spaces is localized when inserted correlators (or the
instanton) collide (DKM compactification) and reduces to the
multiplication of the lower amplitudes as is typical in the
topological gravity.

On the other hand, the duality between the $(4n+2)$- (or $n$-)
punctured spheres and the genus expansion is reminiscent of the
relation between the conventional David-Distler-Kawai (DDK)
approach \DavidHJ\ and the Liouville F-model approach
to the noncritical string theory. In our SYM case, the genus
expansion is mapped to the tree-level amplitude (but in the
nontrivial background: the insertion of ${1\over a^4} \int d^2z
O(z)$ denotes the deformed background), which reminds us of the
heterotic-type II string duality.\foot{ Another interesting
example of this duality is given by the NS-R duality in the
$\hat{c}=1$ noncritical string theory \GrossZZ. } This must be
another example of the reduction mechanism of the higher genus
amplitude to the punctured sphere studied in section 3.
Furthermore, this analogy enables us to conjecture what
corresponds to the graviphoton correction in the noncritical
string (genus expansion) formalism. In the punctured sphere
approach, we expect that the graviphoton correction comes from the
higher genus amplitude, so in the genus expansion approach, we
expect that the graviphoton correction comes from the perturbation
of the CFT. This should be compared with the `quantum Riemann
surface' approach. It would be very interesting to verify this
conjecture, and for this purpose, the graviphoton corrected
recursion relation (which should be deeply connected with the
higher genus uniformization theory) must be the key step.

At the same time, if we turn the argument around, we have a deep
clue to understand the structure of the path integration over the
quantum Liouville theory. As we have emphasized, our recursion
relation is in the same universality class with the $c=0$
noncritical string theory. Whereas the matrix model solution is
known, it is not yet known how to perform the Liouville path
integration on the higher genus Riemann surfaces, nor how to
integrate over the moduli space in the continuum approach. Since
the direct SYM instanton calculation has been performed recently
and we have obtained the reduction mechanism in this case, we
expect that the similar reduction {\it should} exist in the case
of the $c=0$ noncritical string theory. We believe that a detailed
comparison between direct instanton calculation with our
integration over the moduli space of punctured spheres will lead
us to interesting results.

\subsec{Connection with Gromov-Witten Theory}

It is well-known that the genus zero Gromov-Witten theory on the
local Hirzebruch surface in the geometric engineering limit
determines the prepotential of $SU(2)$ $\N=2$ SYM theory. Thus it
is natural to discuss the connection between our recursion
relation and the underlying recursion relation for the
Gromov-Witten invariants. Such a recursion relation is derived in
\KontsevichQZ\ by using the underlying WDVV equation. For example,
the Gromov-Witten invariants for ${\PP}^2$ satisfy
\eqn\grw{N_d=\sum_{k+l=d}N_kN_lk^2l\left[
l{{3d-4}\choose{3k-2}}- k{{3d-4}\choose{3k-1}}\right] ,} for
$d\ge2$, which resembles our bilinear recursion relation in its
structure.\foot{In \DiFrancescoXR, it was noticed that the
recursion relation for the $N_d$ might be related to $c=0$ two
dimensional gravity. Actually, the nonperturbative differential
equation satisfied by the generating function of $N_d$ is given
by the Painlev\'e VI equation. We also note that this model also
can be regarded as a Liouville F-model by using the same technique
reviewed in section 3.}

To make the physical picture more transparent, take the recursion
relation of the Gromov-Witten invariants for ${\PP}^1 \times
{\PP}^1$ which is the Hirzebruch surface $F_0$:\foot{{The
Gromov-Witten invariants for the local ${\PP}^1 \times {\PP}^1$
are {\it different} from the one for the ${\PP}^1 \times
{\PP}^1$}. The connection between them is not clear yet, but it
should exist as our approach suggests.}
$$N_{n,m}=$$
\eqn\grwt{
\sum_{{n_1+n_2=n}\atop{m_1+m_2=m}}N_{n_1,m_1}N_{n_2,m_2}
(n_1m_2+n_2m_1)m_2\left[ n_1{{2n+2m-4}\choose{2n_1+2m_1-2}}-
n_2{{2n+2m-4}\choose{2n_1+2m_1-1}}\right]\ .} This recursion
relation seems more complicated than ours, which is somewhat
related to the fact that we have double expansions --- $n$ and
$m$. To relate this recursion relation to ours, we first suppose
that a similar recursion relation exists for the local Hirzebruch
surface, which has a direct relation to the SW theory. Then in
order to derive the SW theory, we need a geometric engineering
limit \geomlimit. In this procedure we take a nontrivial
resummation, and the dependence of two parameter reduces to that
of one. For the local Hirzebruch surface, in order to reproduce
the SW solution, the asymptotic Gromov-Witten invariants $d_{n,m}$
for large $m$ must be given by \KatzFH \eqn\asymGW{ d_{n,m} \sim
\gamma_n m^{4n-3}\ ,} where $\gamma_n$ is related to the instanton
amplitudes $\F_n$ as \eqn\asymGWn{\gamma_n = {2^{5n-2} \over
(4n-3)!} 2\pi i \F_n \ .} {}From our bilinear recursion relation,
we can predict the asymptotic recursion relation of the
Gromov-Witten invariants: \eqn\asymGWp{(4n-4)! \gamma_n =
{2^2\over n} \sum_{k=1}^{n-1} g_{k,n} k(n-k) (4k-4)!(4(n-k)-4)!
\gamma_k\gamma_{n-k} \ .} It is very plausible that our bilinear
recursion relation is the remnant of the recursion relation for
the Gromov-Witten invariants. It would be an interesting problem
to see whether we can directly obtain our recursion relation in
this approach. It is also worth mentioning that our construction
of the instanton amplitudes in terms of the rational intersection
theory on $\overline{\M}_{0,n}$ yields the asymptotic growth of
the Gromov-Witten invariants for the local Hirzebruch surface.
This fact indicates that the asymptotic form of the Gromov-Witten
invariants is described by the rational intersection theory on
$\overline{\M}_{0,n}$. We believe some interesting mathematical
structure might be hidden here.

\subsec{Graviphoton Correction}

We have seen that the SW solution in the flat background is deeply
connected to the uniformization theory of the punctured sphere. In
this subsection, we would like to discuss its possible extension
to the graviphoton background. The full prepotential under the
graviphoton background has been proposed recently by Nekrasov
\NekrasovQD\NekrasovRJ. We may also calculate the graviphoton
background prepotential from the geometric engineering approach as
in section 7.1. The direct evaluation of the higher genus
Gromov-Witten invariants is hopeless unless we use the mirror symmetry,
but fortunately, by using the
geometric transition method, the prepotential is calculated as the
Chern-Simons theory, which is now refined as the topological
vertex method \AganagicDB.

Another interesting ingredient in the graviphoton background
$\N=2$ SYM theory is the recent discovery by \FlumeRP\ that if we
properly redefine the modulus $u = {\rm Tr}\,{\tilde{\phi}}^2$,
the $\F$-$u$ relation remains the same even in presence of nontrivial
graviphoton background. This result would suggested considering
the graviphoton corrected PF equations. One may expect bilinear
recursion relations that should also shed lights on the underlying
algebraic geometrical structure.

Understanding the monodromy properties of the corrected
prepotential should lead to a generalization of the uniformization
picture observed in SW theory \MatoneRX\BonelliRY. The relevance
of the uniformization theory in the $\N= 2$ SYM theory is
two-fold. We first observe that any inverse of the uniformization
map $\tau(u) = J^{-1}(u)$ defines a physical coupling constant in
the moduli space $u$ with a suitable monodromy. This is because
$\tau$ is a univalent function, and is defined on the upper-half
plane.\foot{The fact that positivity of the coupling constant may
lead to univalence of coupling constants for effective theories
seems to be at the origin of the underlying stringy nature.
Actually, the inverse of the uniformizing map has a basic property
in the theory of Riemann surfaces. This is strictly related to the
universal Teichm\"uller space $T(1)$ where a nonperturbative
formulation of string theory should be formulated.} Actually, we
can derive the SW solution from the uniformization theory of the
thrice punctured sphere. The second one is related to the
Liouville F-models which we have heavily used to understand the
origin of the genus zero recursion relation.

Let us begin with the geometric engineering setup we have proposed
in section 7.1. When we consider the higher genus (hence
graviphoton) corrections to the prepotential, we naturally expect
from our construction that the free-energy should behave as
\eqn\grap{\F_n^{(g)} \sim \int_{\overline{\cal M}_{g,n+2}}
{\omega_{g,n+2}}^{n-2} \wedge \omega_I \sim\left\langle
\int_{\overline{{\cal M}}_g} dm \left(\int d^2z O(z) \right)^{n}
\right\rangle_{S_\infty,g} \ .} We have used the assumption that
$O(z)$ is almost BRST exact and the contribution of the correlator
comes only from the boundary of the moduli spaces. This is
guaranteed by the Liouville F-model like construction here and the
Wolpert restriction phenomenon. If we evaluate this integral by
using the similar technique as we have employed in section 7.2
(note that we need higher genus calculations leading to the genus
expansion), we obtain the recursion relation which relates the
higher genus amplitude to the lower genus amplitudes. This is
expected from the topological string theory: we have a holomorphic
anomaly equation \BershadskyCX\ which connects higher genus
amplitude with the lower genus ones on one hand, and on the other
hand we know that the different genus contributions are combined in
the Gopakumar-Vafa invariant form \GopakumarJQ\ before taking the
geometric engineering limit.

The other possibility is that we consider the {\it quantum}
uniformization theory on the thrice punctured Riemann surface.
This nicely fits the existence of the quantum corrected PF
equation. Also, if one considers the geometric engineering and
performs a mirror symmetry, the target Kodaira-Spencer theory in the
mirror B-model is
locally described by that of the Riemann surface. Since quantum
Kodaira-Spencer theory provides the genus effect, the concept of
the quantum uniformization must occur naturally. Our quantum
uniformization theory should involve the `quantum' Liouville
theory in contrast to the classical Liouville theory we have
mainly utilized in this paper. In the quantum Liouville theory,
the quantum correction parameter $b$, which is related to the
central charge as $c= 1+6(b+b^{-1})^2$, is identified with the
Planck constant $\hbar$ which is the graviphoton correction. It
would be an interesting problem to check whether this conjecture
is true or not and what is the actual uniformization geometry.

There exists an interesting possibility that the recent progress
in string and SYM theories may lead to a deeper understanding of
the uniformization theory, in particular on the theory of modular
functions. To see this, let us first consider the genus one
correction to the SW prepotential. This is given in terms of the
$\eta$ function \KlemmPA\ \eqn\geereuno{ \F^{(1)}=-\ln \eta(\tau)
\ . } Now note that the $\eta$ function has a well-defined
$SL(2,\ZZ)$ monodromy. On the other hand, such a monodromy is the
uniformizing group of the sphere with three singularities: one
puncture at $\infty$ and elliptic fixed points of orders 2 and 3.
Thus we see that whereas the genus zero prepotential is related to
the thrice punctured sphere, in the case of the genus one
contribution there appears the uniformizing group for the sphere
with elliptic points.\foot{We note that this reflects the
monodromy group. In particular, whereas in the case of the thrice
punctured sphere the second derivative of the prepotential with
respect to $a$ is the inverse of the uniformizing map, in genus
one there is still a well defined monodromy, that now is
$SL(2,\ZZ)$, for the prepotential. In fact the relation between
the uniformizing map and the prepotential is different with
respect to the case of genus zero.}

What about the underlying geometry of the higher genus
contributions? There are some interesting suggestions. First, we
saw that the instanton moduli space is strictly related to the
geometry of $\overline\M_{0,n}$ and of the DKM stable
compactification. On the other hand, we saw that there is a
natural mapping between Hurwitz spaces and $\overline\M_{0,n}$. It
should be also noted that if the higher genus contributions were
related to the uniformization of Riemann surfaces, this would
provide a tool to obtain exact results. However, the
uniformization theory is a long-standing problem and hense one
should expect that new insights should involve particular
situations although not yet discovered.

A trivial example in which the uniformization has been solved is
when the punctures are at the root of the unity: the constraints
on the accessory parameters are sufficient to fix them. In this
context, the possible geometry related to the higher genus
contributions to the prepotential, may be the one of surfaces which
are branched covering. Such surfaces have a high symmetry so it
may happen that they are `exactly solvable'. Furthermore,
recently it has been shown that in the case of branched covering
of the torus one may obtain explicit solutions for the
eigenfunctions of the Laplacian \MatoneUY. Therefore, even if
uniformization and spectra on Riemann surfaces are apparently
technically very difficult to solve, there are highly symmetric
cases in which these problems have been understood. On the other
hand, there are surfaces which are coverings of lower genus
Riemann surfaces which arise in matrix model theory, see for
example \EynardXT. In the case of the branched covering of the
torus, the high symmetry of these particular surfaces reflects in
a sort of Dirac constraint of the Riemann period matrix
$\Omega_{jk}$ \MatoneTJ\MatoneUY\foot{It would be very interesting
to understand the Schottky problem for such period matrices.
Presumably these number theoretical conditions, which leads to
complex multiplication (CM)
for the Jacobian \MatoneUY, should imply some identity related to
the Fay trisecant identity (see for example \MatoneKR). This would
be of interest in considering the moduli integration in string
theory, where the Fay trisecant identity, as first observed by
Eguchi and Ooguri \EguchiUI, corresponds to the bosonization
formula.}
\eqn\bellissima{m_j'-\sum_{k=1}^g\Omega_{jk}n_k'=\overline c
\left(m_j-\sum_{k=1}^g\Omega_{jk}n_k\right)\ , } where
$m_j,n_j,m_j',n_j'$ are integers. Remarkably, this condition
naturally appears in string theory \GrignaniZM, suggesting that
more generally there is a selection of the geometry contributing
to the genus expansion. It is interesting to observe that the
above structures are related to the properties of the Abelian
differentials and to the theory of quantum billiards
\Kontsevichetal.

Since the basic underlying group for the prepotential is
$SL(2,\ZZ)$, and considering that in genus zero we have the
uniformizing group $\Gamma(2)\subset SL(2,\ZZ)$, one should
investigate whether the higher symmetry we discussed reflects in
fixing some specific subgroups of $SL(2,\ZZ)$ as the monodromy
groups of the higher genus contributions to the prepotential.
Therefore we should have the sequence of monodromy (uniformizing)
groups \eqn\sequenza{ \Gamma_0=\Gamma(2)\subset SL(2,\ZZ) \qquad
\Gamma_1=SL(2,\ZZ) \qquad \Gamma_2\subset SL(2,\ZZ) \qquad
\Gamma_3\subset SL(2,\ZZ) \ \ldots \ , } with the generic
$\Gamma_g$ subgroup of $SL(2,\ZZ)$.

There is a related intriguing structure which needs to be
mentioned. Namely, the thrice punctured sphere in $\N=2$ instanton
theory with gauge group $SU(2)$ also appears for the {\it same}
group and in the {\it same} context, but now for the classical
theory. More precisely, it turns out \LTA\ that the self-dual
Yang-Mills (SDYM) equation \eqn\nnU{F=*F \ ,} for the gauge group
given by the volume preserving diffeomorphisms of $SU(2)$, has
solutions parametrized just by the uniformizing equation for the
thrice punctured sphere! Remarkably, there are two interesting
solutions, one is described by the Hauptmodule, i.e. inverse of
the uniformizing map, for $H/SL(2,\ZZ)$, while the other concerns
$H/\Gamma(2)$, exactly the first two groups arising in the genus
zero and one contribution to the prepotential.\foot{In \LTA\ it
has been obtained a differential equation relating the two
Hauptmodules based on the well known relation. These relations
imply a connection between the genus zero and genus one
contributions to the prepotential. It would be interesting to
study such equations to gain insights on the existing relations
between different genus contributions to the prepotential. This
would relate the monodromies.} Interestingly, while the genus one
prepotential in $\N=2$ SYM concerns the graviphoton corrections,
the self-dual reduction considered in \LTA\ concerns the Bianchi
IX cosmological model (see also \HarnadHH\ for the appearence of such
uniformizing equations in other related contexts). On the other hand the appearance of volume
preserving diffeomorphisms of $SU(2)$ just indicates the emergence
of gravity. We believe that this promising analogy deserves to be
further understood. A first interesting question is to understand
whether the generalization to higher rank groups of such
`gravitational SDYM equations' are in turn similarly related to
the higher rank PF equations of SW theory.

\subsec{Higher Rank Gauge Groups}

Here we would like to discuss another possible extension of our
results, namely to the higher rank gauge groups \KlemmQS. First of
all, we note that the general expression corresponding to the
$\F$-$u$ relation \mato\ for the higher rank $\N=2$ SYM theory was
derived in \SonnenscheinHV. This relation should be the basis of
our construction. We also note that similar relations exist
between the higher rank group invariants \BonelliQH\DHokerJM.

Our construction of the $SU(2)$ gauge group is based on the
Liouville F-models, which provides a certain universality class of
the string theory including the $c=0$ quantum gravity. Actually,
we have pointed out many similarities between the $SU(2)$ SW
theory and the $c=0$ noncritical string theory in this paper.
Therefore it is natural to relate the higher rank gauge groups SW
theory to the ADE minimal models coupled to the two-dimensional
quantum gravity (Liouville theory). In particular, $SU(N)$ gauge
theory is related to the $A_{N-1}$ minimal models. Of course,
$SU(2)$ is given by the $A_1$ minimal model coupled to the
gravity, or $c=0$ Liouville theory itself.

The emergence of the ADE minimal models is also expected from the
geometric engineering approach. In the case of the ADE gauge
group, the fiber of the base ${\PP}^1$ is given by the surface
which possesses the ADE singularity in the geometric engineering
limit. At the same time, it is well-known that the ADE singularity
in its local form is described by the ADE $\N=2$ minimal models.
If we twist the theory, $\N=2$ minimal model is supposed to become
the {\it bosonic} (or topological) minimal model, which we couple
to the two-dimensional (topological) gravity.

{}From the duality to the matrix model, we can derive the
nonlinear recursion relation for the free-energy of the unitary
minimal models coupled to the two-dimensional gravity. The
deformation parameter there should be related to the moduli
parameter of the $\N=2$ SYM theory. As in the $SU(2)$ case which
we have thoroughly studied in this paper, we will be able to
obtain the recursion relation for the SW theory by deforming the
underlying recursion relation for the ADE minimal gravity.

Finally we should add that in \MatoneNF, it was realized that the
unitary minimal models coupled to the two-dimensional gravity can
be formulated in a purely geometrical manner. Therefore, there is
a possibility that the $\N=2$ SYM theory with an arbitrary gauge
group can be described solely in terms of the quantum geometry of
the Riemann surface. This would be an interesting problem worth
studying further.

\newsec{Conclusion}

In this paper, we have studied the Liouville geometry of the
$\N=2$, $SU(2)$ SYM theory and proposed a bilinear recursion
relation based on the observation that the theory has
a similar structure with the Liouville geometry of the punctured
sphere. By utilizing the underlying Liouville theory, we succeeded
in presenting the physical origin of the bilinear recursion
relation in three different ways.

While these expressions have a firm ground from the macroscopic
point of view, it remains a very interesting problem to derive
them from the more microscopic point of view. Let us state three
major microscopic problems here to conclude the whole paper and
indicate the future directions.

Firstly, from our study we believe in the existence of the direct
map between the ADHM moduli space and the moduli space of the
punctured spheres, and the connection between the various
localization formula and our boundary of the moduli space. Thus
the rederivation of our results from the direct instanton
calculation along the recent developments around Nekrasov's
complete solution should cast a new insight into the structure of
the instanton moduli space. Secondly, in the geometric engineering
approach, the derivation of the bilinear recursion relation as a
consequence of the underlying relation for the Gromov-Witten
invariants, whose possibility we have sketched in section 8.2 is
another intriguing challenge from the microscopic perspective.
Finally, in the noncritical string approach, the central problem
is to find a microscopic action for the theory, whose discovery
and understanding of its reduction mechanism should also bring a
new impact on the evaluation of the Liouville path integral from
the first principle. We would like to report the progress on these
fascinating subjects in the near future.

\bigskip

\noindent{\bf Acknowledgements:} M.~M. acknowledges F. Fucito for
interesting discussions and S.~A. Wolpert for illuminating
comments on the restriction phenomenon, Y.~N. acknowledges
K.~Sakai and Y.~Tachikawa for valuable discussions. G.~B. is
supported by the Foundation BLANCEFLOR Boncompagni-Ludovisi,
n\'{e}eBuildt. M.~M. and L.~M. are partially supported by the
European Community's Human Potential Programme under contract
HPRN-CT-2000-00131 Quantum Spacetime.

\vfill\eject

\appendix{A}{The Restriction Phenomenon and the Weil-Petersson Divisor}

\subsec{\it Wolpert's theorem}

Let us start by proving the restriction phenomenon. Namely we show
that from the natural embedding
$$
i: \overline{V}^{(m)}\to
\overline{V}^{(m)}\times * \to
\overline{V}^{(m)}\times
\overline{V}^{(n-m+2)}
\to \partial \overline{V}^{(n)}
\to \overline{V}^{(n)}, \qquad n>m\ ,
$$
where $*$ is an arbitrary point in $\overline{V}^{(n-m+2)}$, it
follows that \wolpertis \eqn\appa{ \left[\omega_{m}\right]=
i^*\left[\omega_{n}\right], \qquad n>m\ .} In order to prove
\appa\ we need to consider the Fenchel-Nielsen parametrization of
the Teichm\"uller space (see for example \bookss). Let $\{P_i\}$
be a set of surfaces homeomorphic to $\widehat {\CC}$ minus three
open discs. Each $P_i$ has a hyperbolic structure with geodesic
boundary whose length may be arbitrarily prescribed in the
interval $[0,\infty)$ (a length $0$ corresponds to a puncture).
Let \eqn\sigmama{ \Sigma_{0,n}={\CC}\backslash\{z_1,\ldots ,
z_{n-3},0,1\}\ ,} be a genus $0$ surface with $n$ punctures. It
can be obtained by glueing $\{P_i\}_{i=1,\dots,n-2}$ identifying
the different boundary components in $n-3$ geodesics on
$\Sigma_{0,n}$. Clearly, to completely characterize the glueing
procedure, we need also to distinguish twisted boundary
components. To this end, for each geodesic $\alpha_i$ we denote by
$\tau_i$ the coordinate describing twists from an arbitrary
reference position. Denoting by $l_i=l_{\alpha_i}$ the length of
each geodesic, we define the Fenchel-Nielsen form \eqn\FneNie{
\omega_{FN}^{(n)}=\sum_{j=1}^{n-3} dl_j\wedge d\tau_j=
\sum_{j=1}^{n-3}l_j dl_j\wedge d\theta_j\ ,} where $\theta_j$ is
the twisting angle (it has been proved that $\omega_{FN}^{(n)}$
does not depend on the particular geodesic dissection of
$\Sigma_{0,n}$).

The observation that the smooth reduction of $\omega_{FN}^{(n)}$
to $\partial \overline V^{(n)}$ is performed letting one or more
geodesical lengths go to $0$ giving a well-defined geodesical
dissection, implies $[\omega_{FN}^{(m)}]= i^*[\omega_{FN}^{(n)}]$,
$n>m$. Eq.\appa\ follows by noticing that \wolpertis \eqn\restwp{
[\omega_{FN}^{(n)}]=[\omega_{n}]\ ,} in $H^2\left(\overline
V^{(n)},{\RR}\right)$.

\subsec{\it Moving Puncture and $D_{WP}^{(n)}$}

Now, following \Zog, we show that \eqn\devi{
D_{WP}^{(n)}={\pi^2\over n-1}\sum_{k=1}^{[n/2]-1}k(n-k-2)D_k \ .}
Observe that the coordinate of a Riemann surface can be seen as a
moving puncture. Therefore, we can consider the embedding of
$\Sigma_{0,n-k}$ in $V^{(n-k+1)}$ \eqn\nembe{
\Sigma_{0,n-k}\longrightarrow V^{(n-k+1)}\ , \qquad z\mapsto
(z_1,\ldots , z_{n-k-3},z)\in V^{(n-k+1)}\ ,\quad z\in
\Sigma_{0,n-k}\ .} Observe that $\Sigma_{0,n-1}$ embeds into
$V^{(n)}$ and therefore also into $\overline V^{(n)}$. A natural
embedding into $\overline V^{(n)}$ can be defined also for the
surfaces
 $\Sigma_{0,n-k}$, $k=2,...,[n/2]-1$, namely
\eqn\appg{\Sigma_{0,n-k}\to V^{(n-k+1)}\to {\overline
V}^{(k+1)}\times {\overline V}^{(n-k+1)}\to D_{k-1}\to {\overline
V}^{(n)}\ ,\quad k=2,\ldots, [n/2]-1\ .} The closure of the image
of $\Sigma_{0,n-k}$ in ${\overline V}^{(n)}$ defines a 2-cycle
$C_k$ isomorphic to $\widehat{\CC}$. By \appa\ and \appg\ it
follows that \eqn\apph{ \left[\omega_{n}\right]\cap
\left[C_k\right]=\int_{i\Sigma_{0,n-k}}\omega_{n}=
\int_{\Sigma_{0,n-k}}i^*\omega_{n}=\int_{\Sigma_{0,n-k}}
\omega_{n-k+1}\ ,} where $\cap$ denotes the topological cap
product. Note that $\left[\omega_{n}\right]\cap
\left[C_k\right]=D_{WP}^{(n)}\cdot C_k$ where $\cdot$ denotes the
topological intersection (see for example \GH). In order to
perform the last integral we use \wp\ and the asymptotic behavior
of the classical Liouville action when the punctures coalesce
\zztop \eqn\apprr{
\partial_{z_i}S^{(n)}_{cl}(z_1,\ldots,z_{n-3})=
\cases{{\pi\over z_i-z_k}+o\left({1\over |z_i-z_k|}\right)\ ;
\quad z_i\to z_k\ ,\quad k\ne n\ ;\cr {\pi\over
z_i}+o\left({1\over |z_i|}\right)\ ;\qquad z_i\to\infty\ .} } Now
observe that\foot{The integrals are understood in the sense of
Lebesgue measure.} \eqn\observ{ D_{WP}^{(n)}\cdot
C_k=\int_{\CC}\omega_{n-k+1}= -{1\over 2i}\lim_{r\to
0}\int_{{\CC}\backslash \Delta_r}
d\partial_{z_{n-k-2}}S^{(n-k+1)}_{cl}dz_{n-k-2}\ ,} where
$\Delta_r$ is the union of $n-k-1$ disks of radius $r$ centered at
$z_1,\ldots,z_{n-k-3},0,1$. Let us now set $z\equiv z_{n-k-2}$.
Since $\partial_z S^{(n-k+1)}_{cl}\in {\cal
C}^\infty\left({\CC}\backslash \Delta_r\right)$, we can apply
Stokes theorem
$$
\int_{{\CC}\backslash \Delta_r}
d\partial_{z}S^{(n-k+1)}_{cl}dz
$$
\eqn\stokes{ =\int_{\partial {\CC}}\partial_{z}S^{(n-k+1)}_{cl}dz-
\int_{\partial \Delta_r}\partial_{z}S^{(n-k+1)}_{cl}dz= 2i\pi^2
-2i\pi^2 (n-k-1)\ .} On the other hand \eqn\limtt{ \lim_{r\to 0}
\int_{\Delta_r}d\partial_{z}S^{(n-k+1)}_{cl}dz=0 \ ,} so that
\eqn\appwe{ D_{WP}^{(n)}\cdot C_k=\pi^2(n-k-2)\ .} Eq.\devi\
follows immediately from \appwe\ and from the non singular matrix
$A_{jk}=C_j\cdot D_k$ of intersection numbers
 between the $2$-cycles $C_j$ and the $(2n-8)$-cycles $D_k$ \Zog
\eqn\appsd{
A=\pmatrix{n-1&0&0&0&\ldots&0&0\cr
   n-4&1&0&0&\ldots&0&0\cr
   n-4&-1&1&0&\ldots&0&0\cr
   n-5&0&-1&1&\ldots&0&0\cr
   \vdots&\vdots&\vdots&\vdots&\vdots&\vdots&\vdots\cr
   n-[n/2]&0&0&0&\ldots&-1&1\cr} \ ,}
where $C_j\cdot D_1=n-j-1$ for $j\ge 4$.

\listrefs

\end